\documentclass{cernyrep}
\usepackage{texnames}
\usepackage[T1]{fontenc}
\usepackage{varwidth}
\usepackage{xcolor}

\begin{document}
\title{Introduction to Landau Damping}

\author{W. Herr}

\institute{CERN, Geneva, Switzerland}

\maketitle 

\begin{abstract}
The mechanism of Landau damping is observed in various systems from plasma
oscillations to accelerators.
Despite its widespread use, some confusion has been created,
partly because of the different mechanisms producing the damping but
also due to the mathematical subtleties treating the effects.
In this article the origin of Landau damping is demonstrated for
the damping of plasma oscillations. In the second part it is
applied to the damping of coherent oscillations in particle
accelerators. The physical origin, the mathematical treatment
leading to the concept of stability diagrams and the applications
are discussed.
\end{abstract}

\section{Introduction and history}
Landau damping is referred to as the damping of a collective mode of oscillations
in plasmas without collisions of charged particles.
These Langmuir~\cite{bib:langmuir} oscillations consist of particles with
long-range interactions and cannot be treated with a simple picture involving
collisions between charged particles.
The damping of such collisionless oscillations was predicted by Landau~\cite{bib:landau1946}.
Landau deduced this effect from a mathematical study without reference to a physical
explanation. Although correct, this derivation is not rigorous from the mathematical
point of view and resulted in conceptual problems.
Many publications and lectures have been devoted to this subject~\cite{bib:bohm1,bib:bohm2,bib:sagan}.
In particular, the search for stationary solutions led to severe problems.
This was solved by Case~\cite{bib:case}  and Van Kampen~\cite{bib:kampen} using normal-mode expansions.
For many theorists Landau's result is counter-intuitive and the mathematical treatment
in many publications led to some controversy and is still debated.
This often makes it difficult to connect mathematical structures to reality.
It took almost 20 years before dedicated experiments were carried out~\cite{bib:malmberg} to
demonstrate successfully the reality of Landau damping.
In practice, Landau damping plays a very significant role in plasma physics and can be applied
to study and control the stability of charged beams in particle
accelerators~\cite{bib:sessler1, bib:sessler2}.

It is the main purpose of this article to present a physical picture together with
some basic mathematical derivations, without touching on some of the subtle problems
related to this phenomenon.

The plan of this article is the following.
First, the Landau damping in plasmas is derived and the physical picture behind
the damping is shown.
In the second part it is shown how the concepts are used to study the stability of
particle beams.
Some emphasis is put on the derivation of stability diagrams and beam transfer functions (BTFs)
and their use to determine the stability.
In an accelerator the decoherence or filamentation of an oscillating beam due to
non-linear fields is often mistaken for Landau damping and
significant confusion in the community of accelerator physicists still persists today.
Nevertheless, it became a standard tool to stabilize particle beams of hadrons.
In this article it is not possible to treat all the possible applications
nor the mathematical subtleties and the references should be consulted.

\section{Landau damping in plasmas}
Initially, Landau damping was derived for the damping of oscillations in plasmas.
In the next section, we shall follow the steps of this derivation in some detail.

\subsection{Plasma oscillations}
\begin{figure}[t]
\begin{center}
\includegraphics[height= 6.50cm,width=  2.50cm, angle=-90]{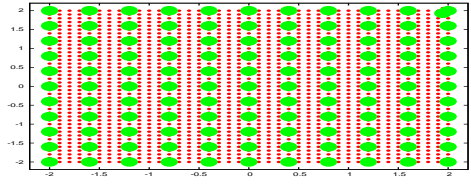}
\end{center}
\caption{Plasma without disturbance (schematic)}
\label{fig:fig1}
\end{figure}

\begin{figure}[t]
\begin{center}
\includegraphics[height= 6.50cm,width=  2.50cm, angle=-90]{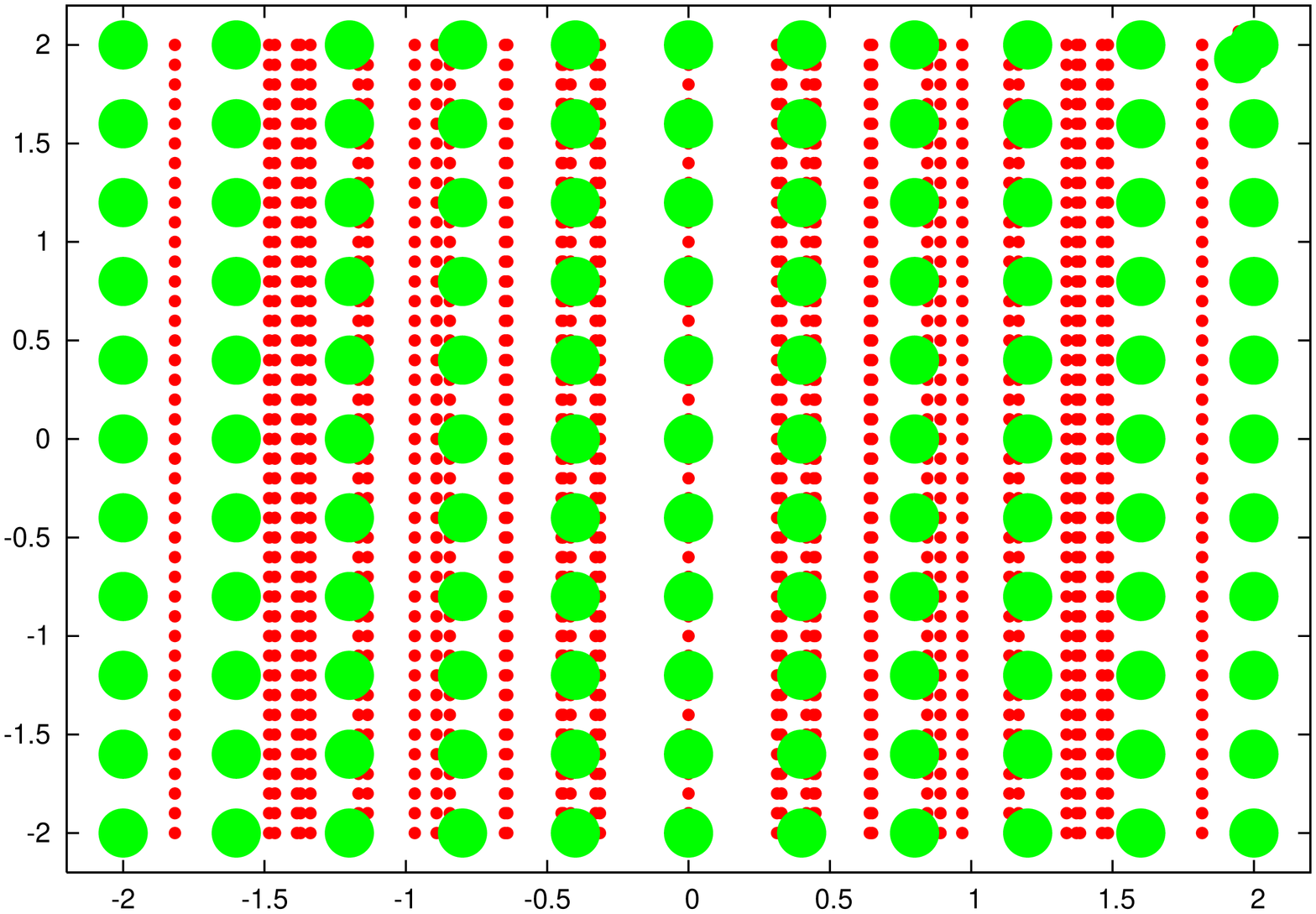}
\end{center}
\caption{Plasma with disturbance (schematic)}
\label{fig:fig2}
\end{figure}

\begin{figure}[t]
\begin{center}
\includegraphics[height= 6.50cm,width=  2.50cm, angle=-90]{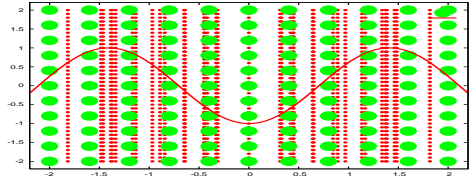}
\end{center}
\caption{Plasma with disturbance and restoring field (schematic)}
\label{fig:fig3}
\end{figure}

We consider an electrically quasi-neutral plasma in equilibrium, consisting of positively
charged ions and negatively charged electrons (\Fref{fig:fig1}).
For a small displacement of the electrons with respect to the ions (\Fref{fig:fig2}), the electric fields act on
the electrons as a restoring force (\Fref{fig:fig3}).
Due to the restoring force, standing density waves are possible with a fixed frequency~\cite{bib:langmuir}
$\omega_{\rm p}^{2} = \frac{n e^{2}}{m \epsilon_{0}}$, where $n$ is the density of electrons,
$e$ the electric charge, $m$ the effective mass of an electron and $\epsilon_{0}$ the permittivity of free space.
The individual motion of the electrons is neglected for this standing wave.
In what follows, we allow for a random motion of the electrons
with a velocity distribution for the equilibrium state and evaluate under which condition
waves with a wave vector ${\bf{k}}$ and a frequency ${\bf{\omega}}$ are possible.

The oscillating electrons produce fields (modes) of the form
\begin{eqnarray}
E(x,t) = E_{0} \sin (kx - \omega t)
\end{eqnarray}
or, rewritten,
\begin{eqnarray}
E(x,t) = E_{0} {\rm e}^{{\rm i}(kx - \omega t)}.
\end{eqnarray}
The corresponding wave (phase) velocity is then $v = \frac{\omega}{k}$.
\begin{figure}[t]
\begin{center}
\includegraphics[height= 6.50cm,width=  2.50cm, angle=-90]{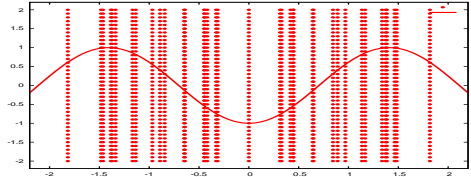}
\end{center}
\caption{Plasma with disturbance}
\label{fig:fig4}
\end{figure}
In \Fref{fig:fig4}, we show the electron distribution together with the produced field.
The positive ions are omitted in this figure and are assumed to produce a stationary, uniform background field.
This assumption is valid when we consider the ions to have infinite mass, which is a good
approximation since the ion mass is much larger than the mass of the oscillating electrons.

\subsection{Particle interaction with modes}
The oscillating electrons now interact with the field they produce, i.e.
individual particles interact with the field produced by all particles.
This in turn changes the behaviour of the particles, which changes the field producing the forces.
Furthermore, the particles may have different velocities.
Therefore, a self-consistent treatment is necessary.
If we allow $\omega$ to be complex ($\omega = \omega_{r} + {\rm i} \omega_{i}$), we separate the real and
imaginary parts of the frequency $\omega$ and rewrite the fields:
\begin{eqnarray}
E(x,t) = E_{0} {\rm e}^{{\rm i}(kx - \omega t)}  \Rightarrow E(x,t) = E_{0} {\rm e}^{{\rm i}(kx - \omega_{\rm r} t)}\cdot {\rm e}^{\omega_{\rm i} t}
\end{eqnarray}
and we have a {\textit{damped}} oscillation for $\omega_{\rm i} < 0$.

If we remember that particles may have different velocities, we can consider a much simplified picture as follows.
\begin{itemize}
\item[i)] If more particles are moving more slowly than the wave:
\item[] Net absorption of energy {\textit{from}} the wave $\rightarrow$ wave is damped.
\item[ii)] If more particles are moving faster than the wave:
\item[] Net absorption of energy {\textit{by}} the wave $\rightarrow$  wave is antidamped.
\end{itemize}
We therefore have to assume that the {\textit{slope}} of the particle distribution at the wave velocity is important.
Although this picture is not completely correct, one can imagine a surfer on a wave in the sea,
getting the energy from the wave (antidamping).
Particles with very different velocities do not interact with the mode and cannot
contribute to the damping or antidamping.

\subsection{Liouville theorem}
The basis for the self-consistent treatment of distribution functions is the Liouville theorem.
It states that the phase-space distribution function is constant along the trajectories of the system,
i.e.\ the density of system points in the vicinity of a given system point travelling
through phase space is constant with time, i.e. the density is always conserved.
If the density distribution function is described by {{$\psi(\vec{q},\vec{p},{{t}})$}},
then the probability to find the system in the phase-space volume ${\rm d}q^{n} {\rm d}p^{n}$ is
defined by $\psi(\vec{q},\vec{p}) {\rm d}q^{n} {\rm d}p^{n}$ with
$ \int {{\psi(\vec{q},\vec{p},{{t}})}} {\mathrm{d}}q^{n} {\mathrm{d}}p^{n} = N$.
We have used canonical coordinates $q_{i}, \ i=1,\ldots,n$ and momenta $p_{i}, \ i=1,\ldots,n$ since
it is defined for a Hamiltonian system.
The evolution in time is described by the Liouville equation:
\begin{eqnarray}
\frac{{\rm d}\psi}{{\rm d}t} = \frac{\partial \psi}{\partial t} + \sum_{i=1}^{n}
\left(\frac{\partial \psi}{\partial q_{i}} {\dot{q_{i}}} + \frac{\partial \psi}{\partial p_{i}} {\dot{p_{i}}}\right) = 0.
\label{eq:e01}
\end{eqnarray}
If the distribution function is {\textit{stationary}} (i.e.\ does not depend on $q$ and $t$), then  {{$\psi(\vec{q},\vec{p},{{t}})$}} becomes {{$\psi(\vec{p})$}}.
\begin{figure}[t]
\begin{center}
\includegraphics*[width=40.8mm,height=40.8mm,angle=-90]{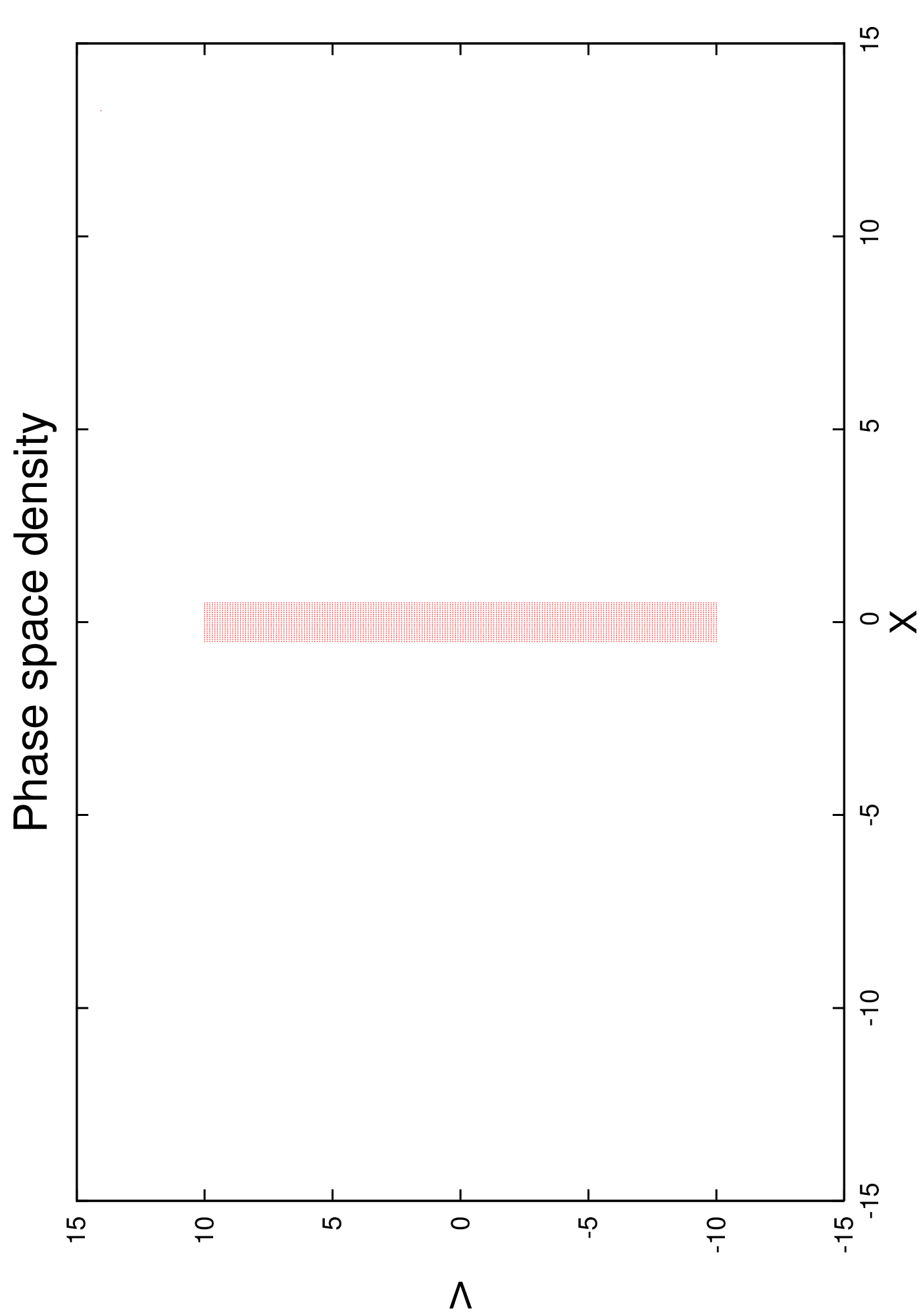}
\includegraphics*[width=40.8mm,height=40.8mm,angle=-90]{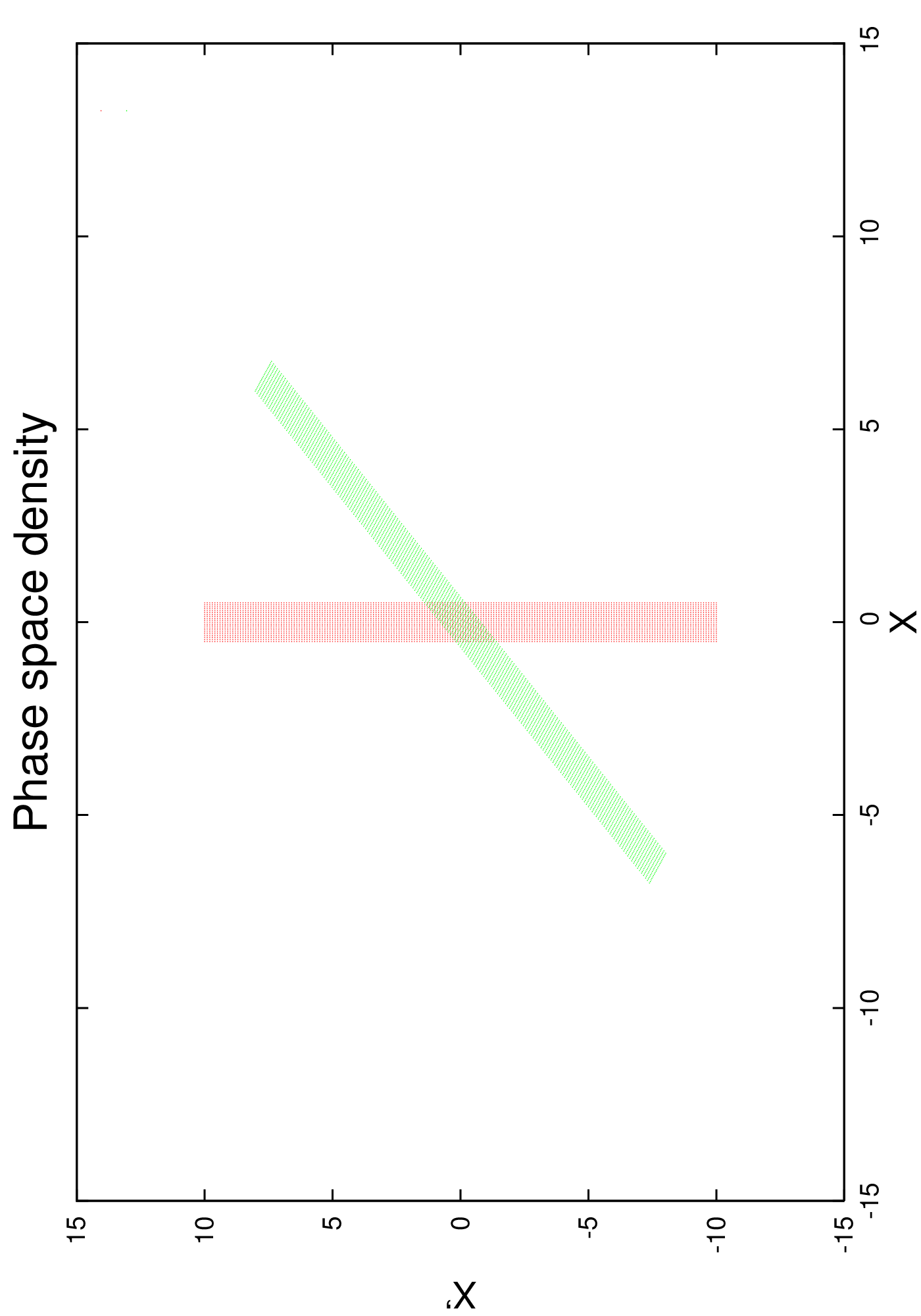}
\includegraphics*[width=40.8mm,height=40.8mm,angle=-90]{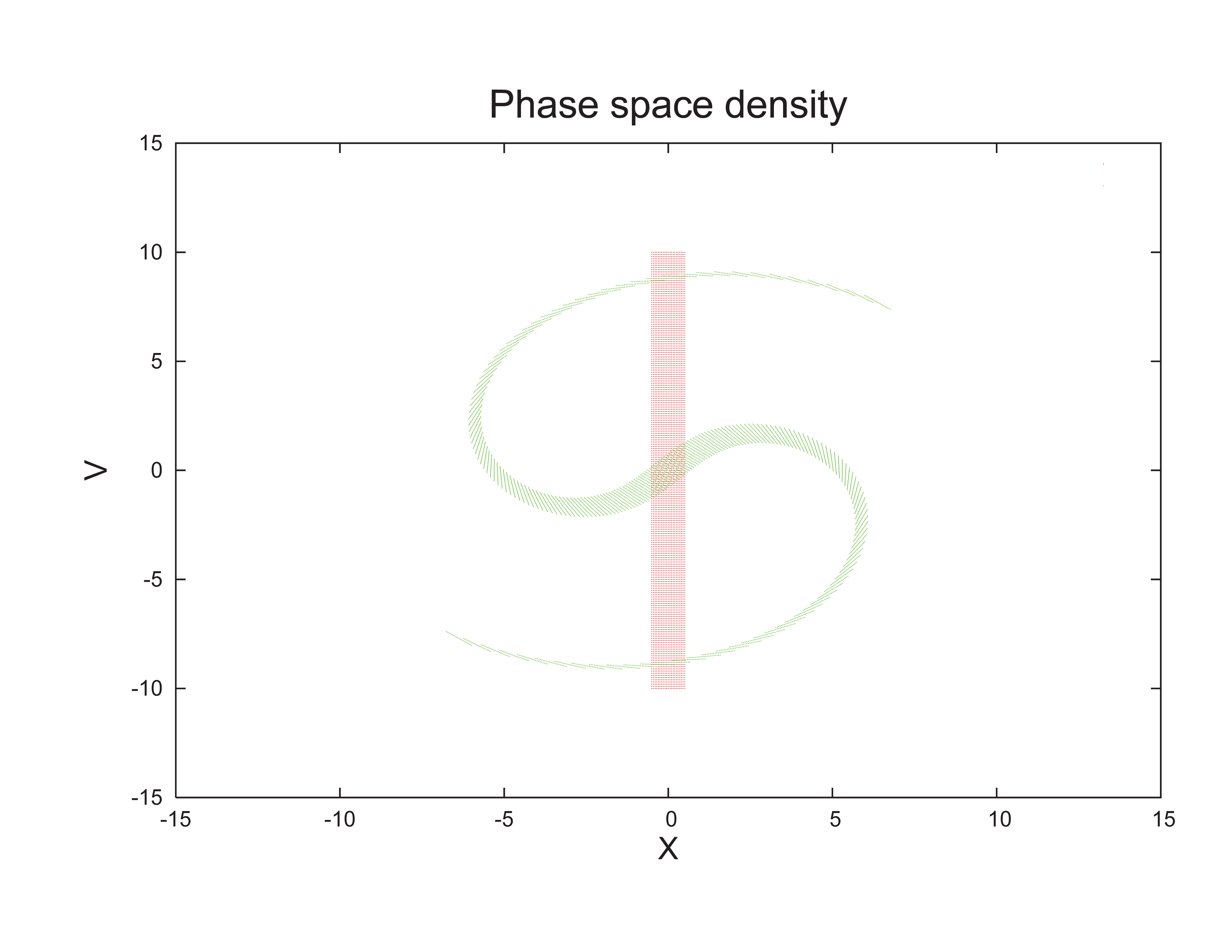}
\end{center}
\caption{Change of phase space. Left, original distribution; centre, distribution with linear fields; right, distribution with non-linear fields after elapsed time $t$.}
\label{fig:fig5}
\end{figure}
Figure 5 shows schematically the change of a phase-space distribution under the influence of a linear and a non-linear field.
The shape of the distribution function is distorted by the non-linearity, but the {\textit{local}} phase-space density is conserved.
However, the {\textit{global}} density may change, i.e. the (projected) measured beam size.
The Liouville equation will lead us to the Boltzmann and Vlasov equations.
We move again to Cartesian coordinates $x$ and $v$.

\subsection{Boltzmann and Vlasov equations}
Time evolution of {{$\psi(\vec{x},\vec{v},{{t}})$}} is described by the Boltzmann equation:
\begin{eqnarray}
\frac{{\rm d}\psi}{{\rm d}t} = \underbrace{\frac{\partial \psi}{\partial t}}_{\mathrm{time~change}}
+ \underbrace{{{\vec{v}}}\cdot\frac{\partial \psi}{\partial \vec{x}}}_{\mathrm{space~change}}
+ \underbrace{{{\frac{1}{m}{\vec{F}(\vec{x},t)}}}\cdot\frac{\partial \psi}{\partial \vec{v}}}_{{{v}~\mathrm{change},~\mathrm{force}~F}}
+ \underbrace{\Omega(\psi)}_{\mathrm{collision}}.
\end{eqnarray}
This equation contains a term which describes mutual collisions of charged particles in the distribution $\Omega(\psi)$.
To study Landau damping, we ignore collisions and the collisionless Boltzmann equation becomes the Vlasov equation:
\begin{eqnarray}
\frac{{\rm d}\psi}{{\rm d}t} = \frac{\partial \psi}{\partial t} + {{\vec{v}}}\cdot\frac{\partial \psi}{\partial \vec{x}}
+ {{\frac{1}{m}{\vec{F}(\vec{x},t)}}}\cdot\frac{\partial \psi}{\partial \vec{v}} = 0.
\end{eqnarray}
Here ${{\vec{F}(\vec{x},t)}}$ is the force of the field (mode) on the particles.

Why is the Vlasov equation useful?
\begin{eqnarray}
\frac{{\rm d}\psi}{{\rm d}t} = \frac{\partial \psi}{\partial t} + {{\vec{v}}}\cdot\frac{\partial \psi}{\partial \vec{x}}
+ {{\frac{1}{m}{\vec{F}(\vec{x},t)}}}\cdot\frac{\partial \psi}{\partial \vec{v}} = 0.
\end{eqnarray}
Here ${{\vec{F}(\vec{x},t)}}$ can be a force caused by impedances, beam--beam effects etc.
From the solution one can determine whether a disturbance is growing (instability, negative imaginary part of frequency) or
decaying (stability, positive imaginary part of frequency).
It is a standard tool to study collective effects.

\subsubsection{Vlasov equation for plasma oscillations}
For our problem we need the force $\vec{F}$ (depending on field $\vec{E}$):
\begin{eqnarray}
    \vec{F} = e \cdot \vec{E}
\end{eqnarray}
and the field $\vec{E}$ (depending on potential ${\Phi}$):
\begin{eqnarray}
    \vec{E} = - \nabla \Phi
\end{eqnarray}
for the potential $\Phi$ (depending on distribution ${\psi}$):
\begin{eqnarray}
    \Delta \Phi = -\frac{\rho}{\epsilon_{0}} = -\frac{e}{\epsilon_{0}} \int \psi \, {\mathrm{d}}v.
\end{eqnarray}
Therefore,  \
\begin{eqnarray}
\frac{{\rm d}\psi}{{\rm d}t} = \frac{\partial \psi}{\partial t} + {{\vec{v}}}\cdot\frac{\partial \psi}{\partial \vec{x}}
+ {{\frac{e}{m}{\vec{E}(\vec{x},t)}}}\cdot\frac{\partial \psi}{\partial \vec{v}} = 0.
\end{eqnarray}
We have obtained a set of coupled equations: the perturbation produces a field which acts back on the perturbation.

Can we find a solution?
Assume a small {\textit{non-stationary}} perturbation {{$\psi_{1}$}} of the {\textit{stationary}} distribution {{$\psi_{0}(\vec{v})$}}:
\begin{eqnarray}
    \psi(\vec{x}, \vec{v}, t) = {{\psi_{0}(\vec{v})}} + {{\psi_{1}(\vec{x}, \vec{v}, t)}}.
\end{eqnarray}
Then we get for the Vlasov equation:
\begin{eqnarray}
\frac{{\rm d}\psi}{{\rm d}t} = \frac{\partial {{\psi_{1}}}}{\partial t} + {{\vec{v}}}\cdot\frac{\partial {{\psi_{1}}}}{\partial \vec{x}}
+ {{\frac{e}{m}{\vec{E}(\vec{x},t)}}}\cdot\frac{\partial {{\psi_{0}}}}{\partial \vec{v}} = 0
\end{eqnarray}
and
\begin{eqnarray}
    \Delta \Phi = -\frac{\rho}{\epsilon_{0}} = -\frac{e}{\epsilon_{0}} \int {{\psi_{1}}}\, {\mathrm{d}}v.
\end{eqnarray}
The density perturbation produces electric fields
which act back and change the density perturbation, which therefore changes with time.
\begin{eqnarray}
    \psi_{1}(\vec{x}, \vec{v}, t) \Longrightarrow {\vec{E}(\vec{x},t)} \Longrightarrow \psi_{1}'(\vec{x}, \vec{v}, t) \Longrightarrow \cdots.
\end{eqnarray}

How can one treat this quantitatively and find a solution for $\psi_{1}$?
We find two different approaches, one due to Vlasov and the other due to Landau.

\subsubsection{Vlasov solution and dispersion relation}
The Vlasov equation is a partial differential equation and we can try to apply standard techniques.
Vlasov expanded the distribution and the potential as a double Fourier transform \cite{bib:vlasov}:
\begin{eqnarray}
    \psi_{1}(\vec{x}, \vec{v}, t) = \frac{1}{2\pi}\int_{-\infty}^{+\infty}\int_{-\infty}^{+\infty}
    {\tilde{\psi_{1}}}(k, \vec{v}, \omega) {\rm e}^{{\rm i}(kx - \omega t)} \, {\mathrm{d}}k\, {\mathrm{d}}\omega,
\end{eqnarray}
\begin{eqnarray}
    \Phi(\vec{x}, \vec{v}, t) = \frac{1}{2\pi}\int_{-\infty}^{+\infty}\int_{-\infty}^{+\infty}
    {\tilde{\Phi}}(k, \vec{v}, \omega) {\rm e}^{{\rm i}(kx - \omega t)}\, {\mathrm{d}}k \,{\mathrm{d}}\omega
\end{eqnarray}
and applied these to the Vlasov equation.
Since we assumed the field (mode) of the form $E(x,t) = E_{0} {\rm e}^{{\rm i}(kx - \omega t)}$,
we obtain the condition (after some algebra)
\begin{eqnarray}
     1 + \frac{e^{2}}{\epsilon_{0} m k} \int \frac{\partial \psi_{0}/\partial v}{(\omega - k v)}\, {\mathrm{d}}v = 0
\end{eqnarray}
or, rewritten using the plasma frequency $\omega_{\rm p}$,
\begin{eqnarray}
     1 + \frac{\omega_{\rm p}^{2}}{k} \int \frac{\partial \psi_{0}/\partial v}{(\omega - k v)}\, {\mathrm{d}}v = 0
\end{eqnarray}
or, slightly re-arranged for later use,
\begin{eqnarray}
     1 + \frac{\omega_{\rm p}^{2}}{k^{2}} \int \frac{\partial \psi_{0}/\partial v}{({\omega}/{k} - v)}\, {\mathrm{d}}v = 0.
     \label{eq:04}
\end{eqnarray}
This is the {{dispersion relation}} for plasma waves, i.e.\ it relates the frequency ${\bf{\omega}}$ with the  wave vector ${\bf{k}}$.
For this relation, waves with frequency ${\bf{\omega}}$ and wave vector ${\bf{k}}$ are possible,
answering the previous question.

Looking at this relation, we find the following properties.
\begin{itemize}
\item[i)] It depends on the (velocity) distribution  $\psi$.
\item[ii)] It depends on the slope of the distribution  $\partial \psi_{0}/\partial v$.
\item[iii)] The effect is strongest for velocities close to the wave velocity, i.e.\  $v \approx {\omega}/{k}$.
\item[iv)] There seems to be a complication (singularity) at  $v \equiv {\omega}/{k}$.
\end{itemize}
Can we deal with this singularity?
Some proposals have been made in the past:
\begin{itemize}
\item[i)] Vlasov's hand-waving argument~\cite{bib:vlasov}: in practice $\omega$ is never real (collisions).
\item[ii)] Optimistic argument~\cite{bib:bohm1,bib:bohm2}:
$\partial \psi_{0}/\partial v = 0$, where $v \equiv \frac{\omega}{k}$.
\item[iii)] Alternative approach~\cite{bib:kampen, bib:case}:
\begin{itemize}
\item[a)] search for stationary solutions (normal-mode expansion);
\item[b)] results in continuous versus discrete modes.
\end{itemize}
\item[iv)] Landau's argument~\cite{bib:landau1946}:
\begin{itemize}
\item[a)] Initial-value problem with perturbation  $\psi_{1}(\vec{x}, \vec{v}, t)$ at $t = 0$ (time-dependent solution with complex $\omega$).
\item[b)] Solution: in time domain use {{Laplace transformation}}; in space domain use {{Fourier transformation}}.
\end{itemize}
\end{itemize}

\subsubsection{Landau's solution and dispersion relation}
Landau recognized the problem as an initial-value problem (in particular for the
initial values $x = 0, v' = 0$) and accordingly used a different
approach, i.e.\ he used a Fourier transform in the space domain:
\begin{eqnarray}
    {\tilde{\psi_{1}}}(k, \vec{v}, t) = \frac{1}{2\pi}\int_{-\infty}^{+\infty} \psi_{1}(\vec{x}, \vec{v}, t) {\rm e}^{{\rm i}(kx)}\, {\mathrm{d}}x,
\end{eqnarray}
\begin{eqnarray}
    {\tilde{E}}(k, t) = \frac{1}{2\pi}\int_{-\infty}^{+\infty} E(\vec{x}, t) {\rm e}^{{\rm i}(kx)}\, {\mathrm{d}}x
\end{eqnarray}
and a Laplace transform in the time domain:\\
\begin{eqnarray}
    \Psi_{1}(k, \vec{v}, p) = \int_{0}^{+\infty} {\tilde{\psi_{1}}}(k, \vec{v}, t) {\rm e}^{(-pt)}\, {\mathrm{d}}t
\end{eqnarray}
\begin{eqnarray}
    {\cal{E}}(k, p) = \int_{0}^{+\infty} {\tilde{E}}(k, t) {\rm e}^{(-pt)}\, {\mathrm{d}}t.
\end{eqnarray}

Inserted into the Vlasov equation and after some algebra (see references), this leads to the modified dispersion relation:
\begin{eqnarray}
     1 + \frac{e^{2}}{\epsilon_{0} m k} \left[ {\rm P.V.}\int \frac{\partial \psi_{0}/\partial v}{(\omega - k v)}\, {\mathrm{d}}v
     {- \frac{{\rm i}\pi}{k}\left( \frac{\partial \psi_{0}}{\partial v}\right)_{v = \omega/k}}\right] = 0
\end{eqnarray}
or, rewritten using the plasma frequency $\omega_{\rm p}$,
\begin{eqnarray}
     1 + \frac{\omega_{\rm p}^{2}}{k} \left[ {\rm P.V.}\int \frac{\partial \psi_{0}/\partial v}{(\omega - k v)}\, {\mathrm{d}}v
     {- \frac{{\rm i}\pi}{k}\left( \frac{\partial \psi_{0}}{\partial v}\right)_{v = \omega/k}}\right] = 0.
     \label{eq:05}
\end{eqnarray}
Here P.V. refers to `Cauchy principal value'.

It must be noted that the second term appears only in Landau's treatment
as a consequence of the initial conditions and is responsible for damping.
The treatment by Vlasov failed to find this term and therefore did not lead to a damping of the plasma.
Evaluating the term
\begin{eqnarray}
     - \frac{{\rm i}\pi}{k}\left( \frac{\partial \psi}{\partial v}\right)_{v = \omega/k},
\end{eqnarray}
We find damping of the oscillations if $\left(\frac{\partial \psi}{\partial v}\right)_{v = \omega/k} < 0$.
This is the condition for Landau damping.
How the dispersion relations for both cases are evaluated is demonstrated in Appendix \ref{sec:app1}.
The Maxwellian velocity distribution is used for this calculation.

\subsection{Damping mechanism in plasmas}
\begin{figure}[t]
\begin{center}
\includegraphics*[width=50.0mm,height=73.0mm,angle=-90]{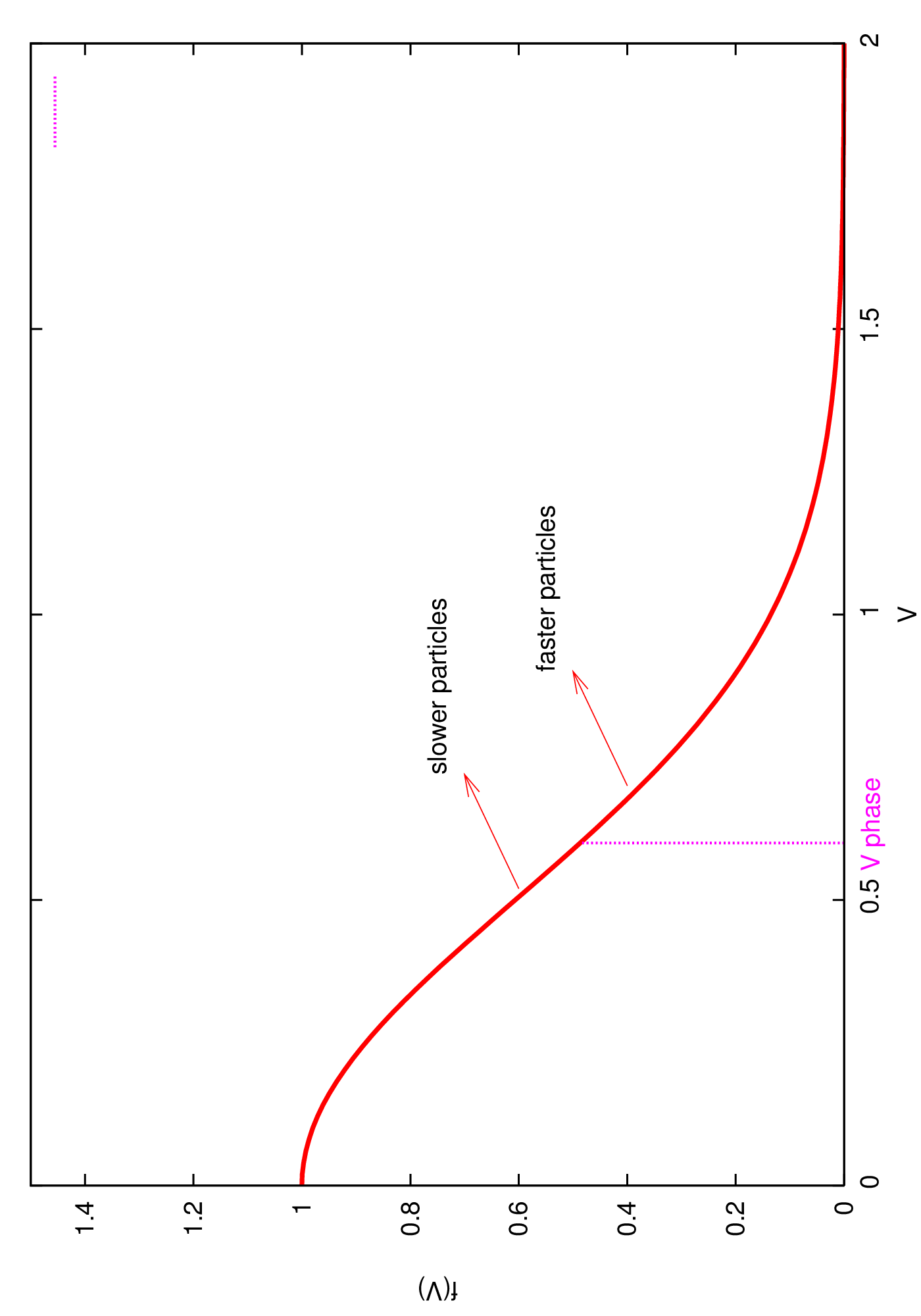}
\end{center}
\caption{Velocity distribution and damping conditions}
\label{fig:fig6}
\end{figure}

Based on these findings, we can give a simplified picture of this condition.
In \Fref{fig:fig6}, we show a velocity distribution (e.g.\ a Maxwellian velocity distribution).
As mentioned earlier, the damping depends on the number of particles below and above the
phase velocity.
\begin{itemize}
\item[i)] More `slower' than `faster' particles $\Longrightarrow$ damping.
\item[ii)] More `faster' than `slower' particles $\Longrightarrow$ antidamping.
\end{itemize}
This intuitive picture reflects the damping condition derived above.

\section{Landau damping in accelerators}
How to apply it to accelerators?
Here we do not have a velocity distribution, but a frequency distribution $\rho(\omega)$ (in the transverse plane the tune).
It should be mentioned here that $\rho(\omega)$ is the distribution of {\textit{external focusing frequencies}}.
Since we deal with a distribution, we can introduce a frequency spread of the distribution and call it $\Delta \omega$.
The problem can be formally solved using the Vlasov equation, but the physical interpretation is very fuzzy (and still debated).
Here we follow a
different (more intuitive) treatment (following \cite{bib:chao, bib:hofmann, bib:sagan}).
Although, if not taken with the necessary care, it can lead to a wrong physical picture, it delivers very
useful concepts and tools for the design and operation of an accelerator.
We consider now the following issues.
\begin{itemize}
\item[i)] Beam response to excitation.
\item[ii)] BTF and stability diagrams.
\item[iii)] Phase mixing.
\item[iv)] Conditions and tools for stabilization and the related problems.
\end{itemize}
This treatment will lead again to a dispersion relation.
However, the stability of a beam is in general not studied by directly
solving this equation, but by introducing the concept of stability diagrams,
which allow us more directly to evaluate the stability of a beam during the design
or operation of an accelerator.

\subsection{Beam response to excitation}
How does a beam respond to an external excitation?

\begin{figure}[t]
\begin{center}
\includegraphics*[width=50.0mm,height=73.0mm,angle=-90]{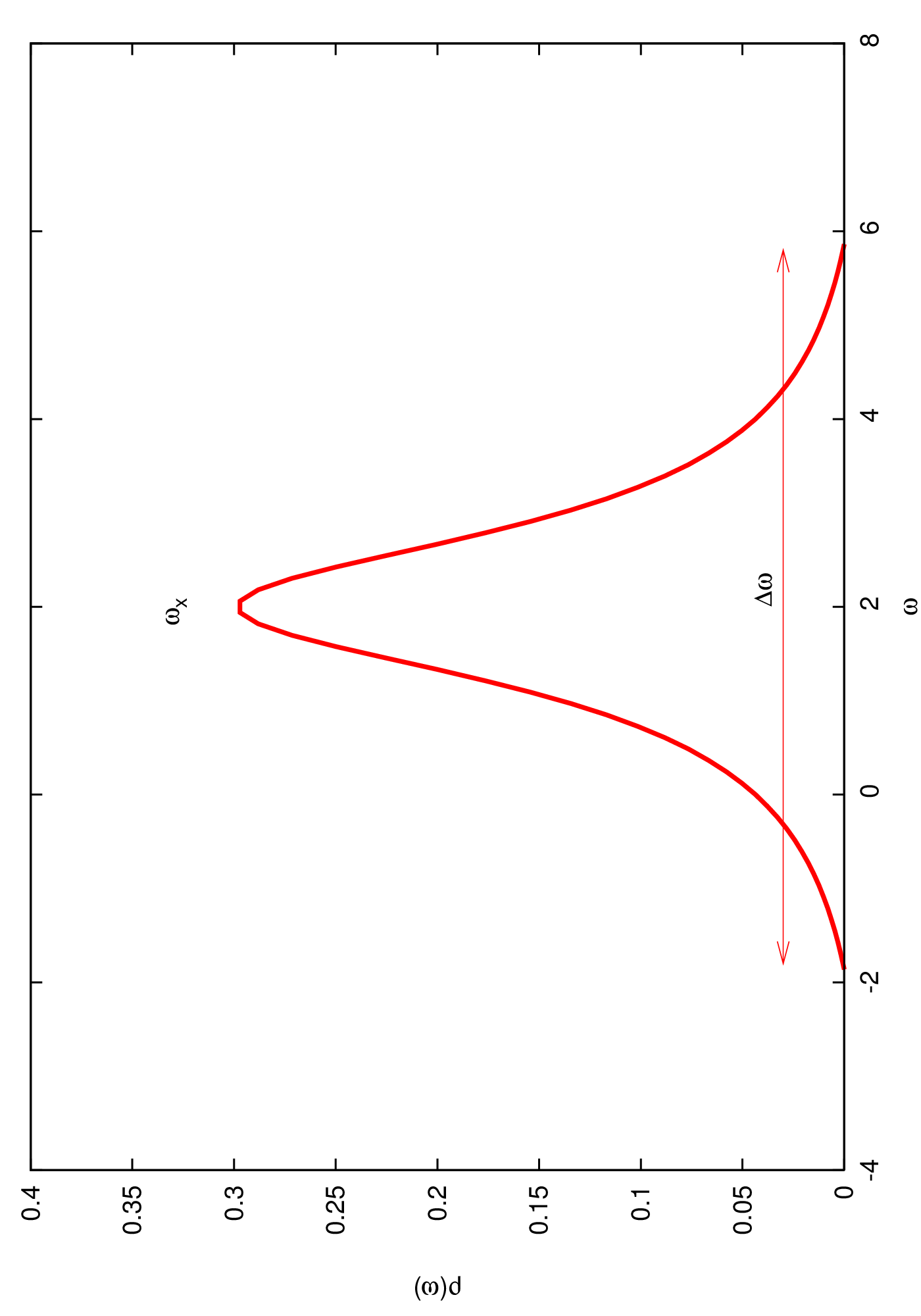}
\end{center}
\caption{Frequency distribution in a beam and frequency spread (very schematic)}
\label{fig:fig7}
\end{figure}

To study the dynamics in accelerators, we replace the velocity $v$ by ${\dot{x}}$ to be consistent with the
standard literature.
Consider a harmonic, {{linear}} oscillator with frequency {{$\omega$}} driven by an external sinusoidal force {{$f(t)$}} with frequency {{$\Omega$}}.
The equation of motion is
\begin{eqnarray}
     \ddot{x} + \omega^{2}x = A \cos \Omega t  = f(t).
\end{eqnarray}
For initial conditions {{$x(0) = 0$}} and {{$\dot{x}(0) = 0$}}, the solution is
\begin{eqnarray}
     x(t) = -\frac{A}{(\Omega^{2} - \omega^{2})} ( \cos \Omega t \underbrace{{{- \cos \omega t}}}_{{{x(0) = 0, \dot{x}(0) = 0}}} ).
\label{eq:06}
\end{eqnarray}
The term ${{- \cos \omega t}}$ is needed to satisfy the initial conditions.
Its importance will become clear later.
The beam consists of an ensemble of oscillators with different frequencies {{$\omega$}} with a
distribution {{$\rho(\omega)$}} and a spread {{$\Delta \omega$}}, schematically shown in \Fref{fig:fig7}.

Recall that for a transverse (betatron) motion {{$\omega_{x}$}} is the {{tune}}.
The number of particles per frequency band is
{{$\rho(\omega) = \frac{1}{N}{\rm d}N/{\mathrm{d}}\omega$}} with {{$\int_{-\infty}^{\infty} \rho(\omega) {\mathrm{d}}\omega = 1$}}.
The {\textit{average}} beam response (centre of mass) is then
\begin{eqnarray}
     \langle x(t)\rangle  = \int_{-\infty}^{\infty} x(t) \rho(\omega) \, {\mathrm{d}}\omega,
\end{eqnarray}
and re-written using (\ref{eq:06})
\begin{eqnarray}
     \langle x(t)\rangle  = - \int_{-\infty}^{\infty}\left[ \frac{A}{(\Omega^{2} - \omega^{2})} ( \cos  \Omega t - \cos  \omega t )\right] \rho(\omega)\, {\mathrm{d}}\omega.
\end{eqnarray}
For a narrow beam spectrum around a frequency $\omega_{x}$ (tune) and the driving force near this frequency ($\Omega \approx \omega_{x}$),
\begin{eqnarray}
     \langle x(t)\rangle  = - \frac{A}{{{2\omega_{x}}}}\int_{-\infty}^{\infty}\left[ \frac{1}{{{(\Omega - \omega)}}} ( \cos  \Omega t - \cos  \omega t )\right] \rho(\omega) {\mathrm{d}}\omega.
\end{eqnarray}
For the further evaluation, we transform variables from {{$\omega$}} to {{$u = \omega - \Omega$}}~(see Ref.\ \cite{bib:chao}) and assume that
{{$\Omega$}} is {{complex}:} {{$\Omega = \Omega_{\rm r} + {\rm i} \Omega_{\rm i}$}} where
\begin{eqnarray}
     \langle x(t)\rangle  & = & - \frac{A}{2\omega_{x}}  \cos(\Omega t) \int_{-\infty}^{\infty} \,{\mathrm{d}}u \rho(u + \Omega) \frac{1 - \cos(u t)}{u}\\
            & ~ & + \frac{A}{2\omega_{x}}  {\sin}(\Omega t) \int_{-\infty}^{\infty} \,{\mathrm{d}}u \rho(u + \Omega) \frac{\sin(u t)}{u}.
\end{eqnarray}
This avoids singularities for $u = 0$.

We are interested in long-term behaviour, i.e.\ $t \rightarrow \infty$, so we use
\begin{eqnarray}
       \lim_{t \rightarrow \infty} \frac{\sin (u t)}{u} = \pi \delta(u),
\end{eqnarray}
\begin{eqnarray}
       \lim_{t \rightarrow \infty} \frac{1 - \cos (u t)}{u} = {\rm P.V.} \left( \frac{1}{u}\right),
\end{eqnarray}
\begin{eqnarray}
       \langle x(t)\rangle  = \frac{A}{2\omega_{x}} \left [ {{\pi \rho(\Omega) \sin(\Omega t)}}
       + {{\cos(\Omega t) {\rm P.V.} \int_{-\infty}^{\infty} {\mathrm{d}}\omega \frac{\rho(\omega)}{(\omega - \Omega)}}} \right].
\end{eqnarray}
This response or BTF has two parts as follows.
\begin{itemize}
\item[i)] {{Resistive}} part, the first term in the expression is in phase with the excitation:
absorbs energy from oscillation $\Longrightarrow$ damping (would not be there without the term {{$- \cos  \omega t$}} in~(\ref{eq:06})).
\item[ii)] {{Reactive}} part, the second term in the expression is out of phase with the excitation.
\end{itemize}
Assuming $\Omega$ is complex, we integrate around the pole and obtain a P.V.\ and a {{`residuum'}} (Sokhotski--Plemelj formula), a standard technique in complex analysis.

We can discuss this expression and find
\begin{itemize}
\item[i)] the `damping' part only appeared because of the initial conditions;
\item[ii)] with other initial conditions, we get additional terms in the beam response;
\item[iii)] that is, for $x(0)~\neq~0$ and $\dot{x}(0)~\neq~0$ we may add
\begin{eqnarray}
 x(0)\int {\mathrm{d}}\omega \,\rho(\omega) \cos(\omega t) + \dot{x}(0)\int {\mathrm{d}}\omega \,\rho(\omega) \frac{\sin(\omega t)}{\omega}.
\end{eqnarray}
\end{itemize}
With these initial conditions, we do not obtain Landau damping and the dynamics is very different.
We have again:
\begin{itemize}
\item[i)] oscillation of particles with different frequencies (tunes);
\item[ii)] now with different initial conditions, {{$x(0) \neq 0$}} and {{$\dot{x}(0) = 0$}} or {{$x(0) = 0$}} and {{$\dot{x}(0) \neq 0$}};
\item[iii)] again we average over particles to obtain the centre of mass motion.
\end{itemize}
We obtain Figs. 8--10.
\begin{figure}[t]
\begin{center}
\includegraphics*[width=50.0mm,height=73.0mm,angle=-90]{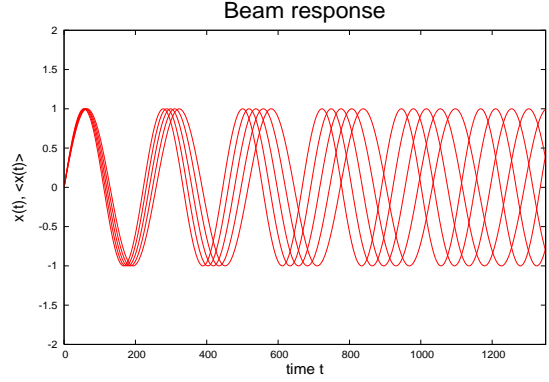}
\end{center}
\caption{Motion of particles with frequency spread. Initial conditions: {{$x(0) = 0$}} and {{$\dot{x}(0) \neq 0$}}}
\label{fig:fig8}
\end{figure}
Figure 8 shows the oscillation of individual particles where all particles have the initial
conditions {{$x(0) = 0$}} and {{$\dot{x}(0) \neq 0$}}.
\begin{figure}[t]
\begin{center}
\includegraphics*[width=50.0mm,height=73.0mm,angle=-90]{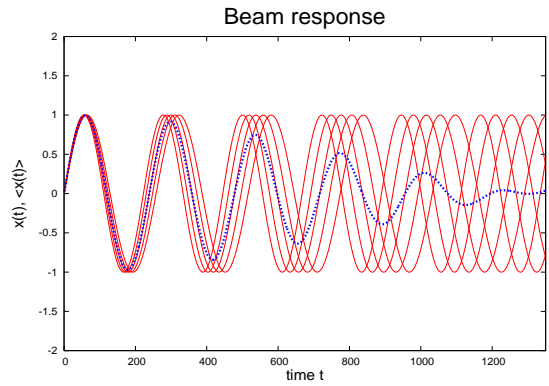}
\end{center}
\caption{Motion of particles with frequency spread and total beam response. Initial conditions: {{$x(0) = 0$}} and {{$\dot{x}(0) \neq 0$}}.}
\label{fig:fig9}
\end{figure}
In \Fref{fig:fig9}, we plot again \Fref{fig:fig8} but add the average beam response.
We observe that although the individual particles continue their oscillations, the average is `damped' to zero.
\begin{figure}[t]
\begin{center}
\includegraphics*[width=50.0mm,height=73.0mm,angle=-90]{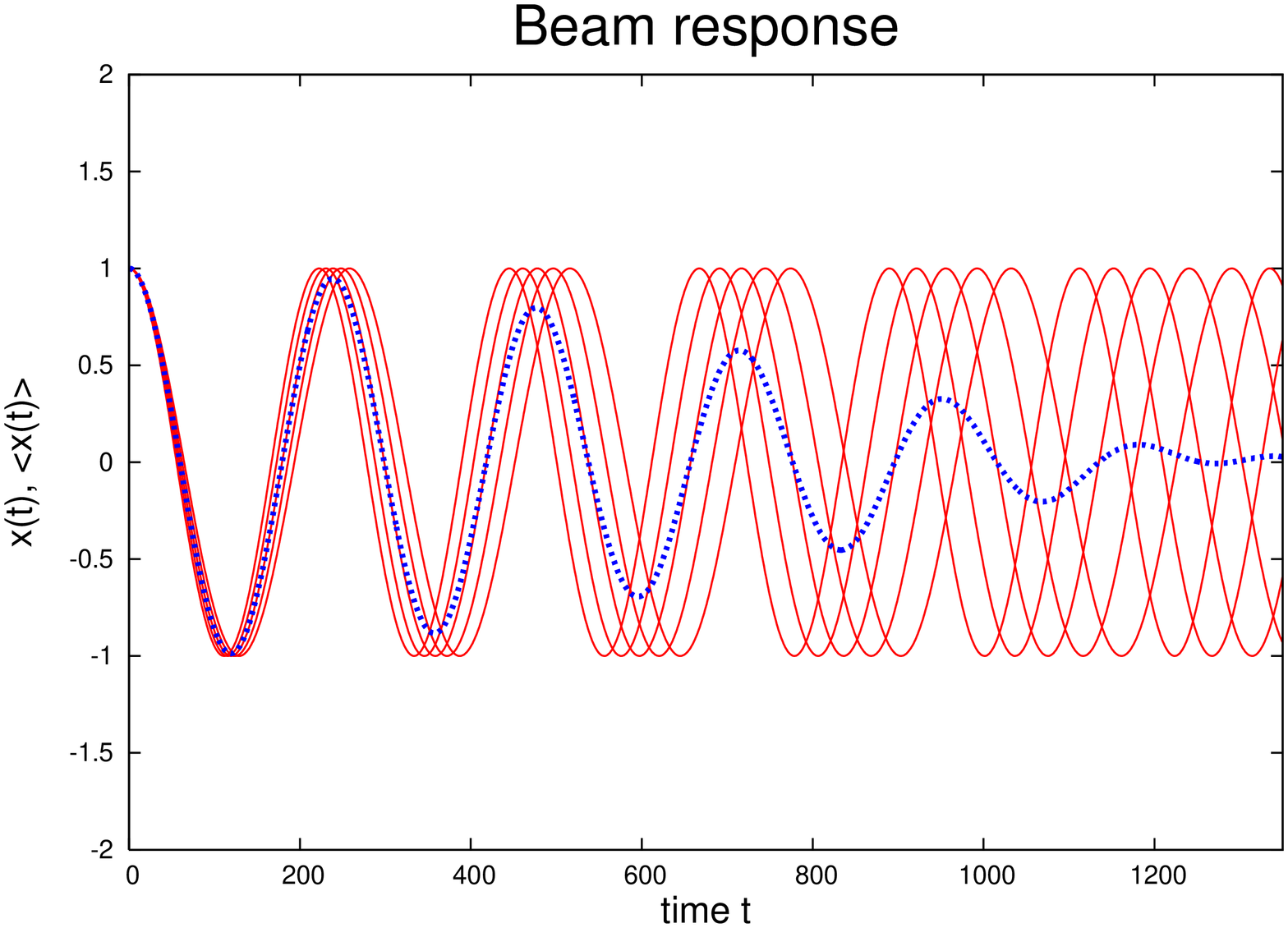}
\end{center}
\caption{Motion of particles with frequency spread and total beam response. Initial conditions {{$x(0) \neq 0$}} and {{$\dot{x}(0) = 0$}}.}
\label{fig:fig11}
\end{figure}
The equivalent for the initial conditions {{$x(0) \neq 0$}} and {{$\dot{x}(0) = 0$}} is shown in \Fref{fig:fig11}.
With a frequency (tune) spread the average motion, which can be detected by a position monitor, damps out.
However, this is \textit{not} Landau damping, rather filamentation or decoherence.
Contrary to Landau damping, it leads to emittance growth.

\subsection{Interpretation of Landau damping}
Compared with the previous case of Landau damping in a plasma, the interpretation of
the mechanism is quite different.
The initial conditions of the beams are important and also the spread of external
focusing frequencies.
For the initial conditions {{$x(0) = 0$}} and {{$\dot{x}(0) = 0$}}, the beam is quiet and
a spread of frequencies {{${{\rho (\omega)}}$}} is present.
When an excitation is applied:
\begin{itemize}
\item[i)] particles cannot organize into collective response (phase mixing);
\item[ii)] average response is zero;
\item[iii)] the beam is kept stable, i.e.\ stabilized.
\end{itemize}
In the case of accelerators the mechanism is therefore not a dissipative
damping but a mechanism for stabilization.
Landau damping should be considered as an `absence of instability'.
In the next step this is discussed quantitatively and the dispersion relations are derived.

\subsection{Dispersion relations}
We rewrite (simplify) the response in complex notation:
\begin{eqnarray}
       \langle x(t)\rangle  = \frac{A}{2\omega_{x}} \left [ {{\pi \rho(\Omega) \sin(\Omega t)}}
       + {{\cos(\Omega t) {\rm P.V.} \int_{-\infty}^{\infty} {\mathrm{d}}\omega \, \frac{\rho(\omega)}{(\omega - \Omega)}}} \right]
\end{eqnarray}
becomes
\begin{eqnarray}
 \langle x(t)\rangle  = \frac{A}{2 \omega_{x}} {{{\rm e}^{-{\rm i}\Omega t}}} \left[ {\rm P.V.} \int {\mathrm{d}}\omega \,
 \frac{\rho(\omega)}{(\omega - \Omega)} + {{{\rm i}\pi \rho(\Omega)}} \right].
\end{eqnarray}
The first part describes an oscillation with complex frequency
${{\Omega}}$.

Since we know that the collective motion is described as {{${\rm e}^{(-{\rm i} \Omega t)}$}},
an {{oscillating solution}} $\Omega$ must fulfil the relation
\begin{eqnarray}
 1 + \frac{1}{2 \omega_{x}} \left[ {\rm P.V.} \int {\mathrm{d}}\omega \frac{\rho(\omega)}{(\omega - \Omega)} + {\rm i}\pi \rho(\Omega) \right] = 0.
\end{eqnarray}
This is again a dispersion relation, i.e.\ the condition for the oscillating solution.
What do we do with that?
We could look where $\Omega_{\rm i} < 0$ provides damping.
Note that no contribution to damping is possible when $\Omega$ is outside the spectrum.
In the following sections, we introduce BTFs and stability diagrams
which allow us to determine the stability of a beam.

\subsection{Normalized parametrization and BTFs}
We can simplify the calculations by moving to {{normalized}} parametrization.
Following Chao's proposal~\cite{bib:chao}, in the expression
\begin{eqnarray}
 \langle x(t)\rangle  = \frac{A}{2 \omega_{x}} {\rm e}^{-{\rm i}\Omega t} \left[ {\rm P.V.} \int {\mathrm{d}}\omega \frac{\rho(\omega)}{(\omega - \Omega)} + {\rm i}\pi \rho(\Omega) \right]
\end{eqnarray}
we use again $u$, but normalized to frequency spread ${\Delta \omega}$.
We have
\begin{eqnarray}
          u =(\omega_{x} - \Omega) \Longrightarrow u = \frac{(\omega_{x} - \Omega)}{\Delta \omega}
\end{eqnarray}
and introduce two functions {{$f(u)$}} and {{$g(u)$}}:
\begin{eqnarray}
          f(u) = \Delta \omega {\rm P.V.} \int {\mathrm{d}}\omega \,\frac{\rho(\omega)}{\omega - \Omega},
     \label{eq:07}
\end{eqnarray}
\begin{eqnarray}
          g(u) = \pi \Delta \omega \rho(\omega_{x} - u\Delta \omega) = \pi \Delta \omega \rho(\Omega).
     \label{eq:08}
\end{eqnarray}

The response with the driving force discussed above is now
\begin{eqnarray}
          \langle x(t)\rangle  = \frac{A}{2\omega_{x} \Delta\omega} {\rm e}^{-{\rm i} \Omega t} \left[f(u) + {\rm i}\cdot g(u) \right],
\end{eqnarray}
where $\Delta\omega$ is the frequency spread of the distribution.
The expression $f(u) + {\rm i}\cdot g(u)$ is the BTF.
With this, it is easier to evaluate the different cases and examples.
For important distributions $\rho(\omega)$ the analytical functions $f(u)$ and $g(u)$ exist (see e.g.~\cite{bib:chaotig})
and will lead us to stability diagrams.

\subsection{Example: response in the presence of wake fields}
\begin{figure}[t]
\begin{center}
\includegraphics*[width=80.0mm,height=23.0mm,angle=-00]{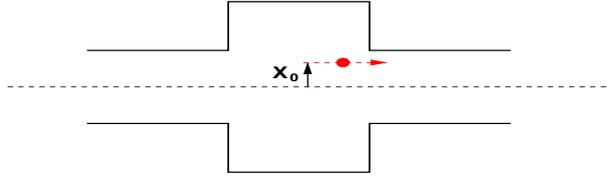}
\end{center}
\caption{Bunch with offset in a cavity-like object}
\label{fig:fig12}
\end{figure}

The driving force comes from the displacement of the beam as a whole, i.e.\ $\langle x\rangle  = X_{0}$, for example driven by a wake field or impedance (see \Fref{fig:fig12}).
The equation of motion for a particle is then something like
\begin{eqnarray}
     \ddot{x} + \omega^{2}x = f(t) = K \cdot \langle x\rangle ,
\end{eqnarray}
where {{$K$}} is a `coupling coefficient'.
The coupling coefficient {{$K$}} depends on the nature of the wake field.
\begin{itemize}
\item[i)] If $K$ is purely real: the force is in phase with the {{displacement}}, e.g. image space charge in perfect conductor.
\item[ii)] For purely imaginary $K$: the force is in phase with the {{velocity}} and out of phase with the displacement.
\item[iii)] In practice, we have both and we can write
\begin{eqnarray}
     K = 2\omega_{x} (U - {\rm i}V).
\end{eqnarray}
\end{itemize}
Interpretation:
\begin{itemize}
\item[a)] a beam travelling off centre through an impedance induces transverse fields;
\item[b)] transverse fields act back on all particles in the beam, via
\begin{eqnarray}
     \ddot{x} + \omega^{2}x = f(t) = K \cdot \langle x\rangle ;
\end{eqnarray}
\item[c)] if the beam moves as a whole (in phase, collectively), this can grow for $V > 0$;
\item[d)] the coherent frequency $\Omega$ becomes complex and is shifted by $(\Omega - \omega_{x})$.
\end{itemize}

\subsubsection{Beam without frequency spread}
For a beam {\textit{without frequency spread}} (i.e. $\rho(\omega) = \delta(\omega - \omega_{x})$), we can easily sum over all particles
and for the centre of mass motion $\langle x\rangle $ we get
\begin{eqnarray}
     \ddot{\langle x\rangle } + \Omega^{2}\langle x\rangle  = f(t) = -2\omega_{x} (U - {\rm i}V) \cdot \langle x\rangle .
\end{eqnarray}
For the original coherent motion with frequency $\Omega$, this means that:
\begin{itemize}
\item[i)] in-phase component $U$ changes the frequency;
\item[ii)] out-of-phase component $V$ creates growth ($V > 0$) or damping ($V < 0$).
\end{itemize}
For any $V > 0$, the beam is unstable (even if very small).

\subsubsection{Beam with frequency spread}
What happens for a beam {\textit{with a frequency spread}}?

The response (and therefore the driving force) was
\begin{eqnarray}
          \langle x(t)\rangle  = \frac{A}{2\omega_{x} \Delta\omega} {\rm e}^{-{\rm i} \Omega t} \left[f(u) + {\rm i}\cdot g(u) \right].
\end{eqnarray}
The (complex) frequency $\Omega$ is now determined by the condition
\begin{eqnarray}
{{-\frac{(\Omega - \omega_{x})}{\Delta \omega}}}  =  {{\frac{1}{(f(u) + {\rm i} g(u))} }}.
\end{eqnarray}
All information about the stability is contained in this relation.
\begin{itemize}
\item[i)] The (complex) frequency difference $(\Omega - \omega_{x})$ contains the effects of impedance, intensity, $\gamma$ etc.
(see the article by G. Rumolo \cite{bib:rumolo}).

\item[ii)] The right-hand side contains information about the frequency spectrum (see definitions for $f(u)$ and $g(u)$ in~(\ref{eq:07}) and~(\ref{eq:08})).
\end{itemize}
Without Landau damping (no frequency spread):
\begin{itemize}
\item[i)] if  ${{\rm Im}}(\Omega - \omega_{x}) < 0$, the beam is stable;
\item[ii)] if  ${{\rm Im}}(\Omega - \omega_{x}) > 0$, the beam is unstable (growth rate $\tau^{-1}$).
\end{itemize}
With a frequency spread, we have a condition for stability (Landau damping)
\begin{eqnarray}
{{-\frac{(\Omega - \omega_{x})}{\Delta \omega}}}  =  {{\frac{1}{(f(u) + {\rm i} g(u))} }}.
\end{eqnarray}
How do we proceed to find the limits?

We could try to find the complex $\Omega$ at the edge of stability  ($\tau^{-1} = 0$).
In the next section, we develop a more powerful tool to assess the stability of a dynamic system.


\section{Stability diagrams}
To study the stability of a particle beam, it is necessary to develop easy to use tools
to relate the condition for stability with the complex tune shift due to, e.g., impedances.
We consider the right-hand side first and call it $D_{1}$.
Both $D_{1}$ and the tune shift are complex and should be analysed in the complex plane.

Using the (real) parameter {{$u$}} in
\begin{eqnarray}
D_{1}  = {{\frac{1}{(f(u) + {\rm i} g(u))} }},
\end{eqnarray}
if we know $f(u)$ and $g(u)$
we can scan $u$ from $-\infty$ to $+\infty$
and plot the real and imaginary parts of $D_{1}$ in a complex plane.

Why is this formulation interesting?
The expression
\begin{eqnarray}
(f(u) + {\rm i} g(u))
\end{eqnarray}
is actually the BTF, i.e.\ it can be {{measured.}}
With its knowledge (more precisely: its inverse), we have conditions on $(\Omega - \omega_{x})$ for stability
and the limits for intensities and impedances.

\subsection{Examples for bunched beams}

As an example, we use a rectangular distribution function for the frequencies (tunes), i.e.
\begin{eqnarray}
\rho(\omega) = \left\{ \begin{array}{ll}
              \frac{1}{2\Delta \omega} & {\mathrm{for}} |\omega - \omega_{x}| \leq \Delta \omega,\\
              0 & {\mathrm{otherwise}}.
              \end{array}
\right.
\end{eqnarray}
We now follow some standard steps.

\medskip

\noindent
Step 1. Compute $f(u)$ and $g(u)$ (or look it up, e.g.\ \cite{bib:chaotig}): we get for the rectangular distribution function
\begin{eqnarray}
f(u) = \frac{1}{2} {\mathrm{ln}}\left| \frac{u + 1}{u - 1}\right|, \quad g(u) = \frac{\pi}{2} \cdot H(1 - \left| u \right|).
\end{eqnarray}

\medskip

\noindent
Step 2. Plot the real and imaginary parts of {{$D_{1}$}}.

\begin{figure}[t]
\begin{center}
\includegraphics*[width=50.0mm,height=73.0mm,angle=-90]{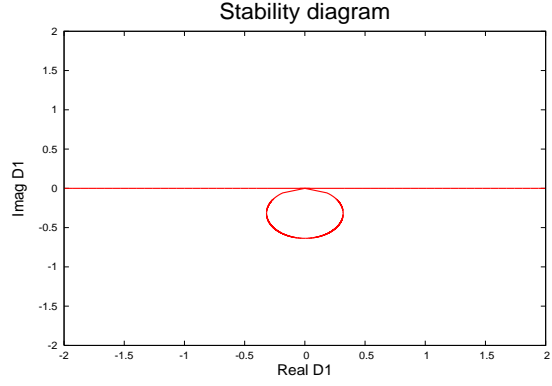}
\end{center}
\caption{Stability diagram for rectangular frequency distribution}
\label{fig:fig13}
\end{figure}
The result of this procedure is shown in \Fref{fig:fig13} and the interpretation is as follows.
\begin{itemize}
\item[$\bullet$] We plot Re$(D_{1})$ versus Im$(D_{1})$ for a rectangular distribution $\rho(\omega)$.
\item[$\bullet$] This is a stability boundary diagram.
\item[$\bullet$] It separates stable from unstable regions (stability limit).
\end{itemize}
Now we plot the complex expression of ${{-\frac{(\Omega - \omega_{x})}{\Delta \omega}}}$ in the same plane as a point (this point depends on impedances, intensities, etc.).
\begin{figure}[t]
\begin{center}
\includegraphics*[width=60.0mm,height=76.0mm,angle=-90]{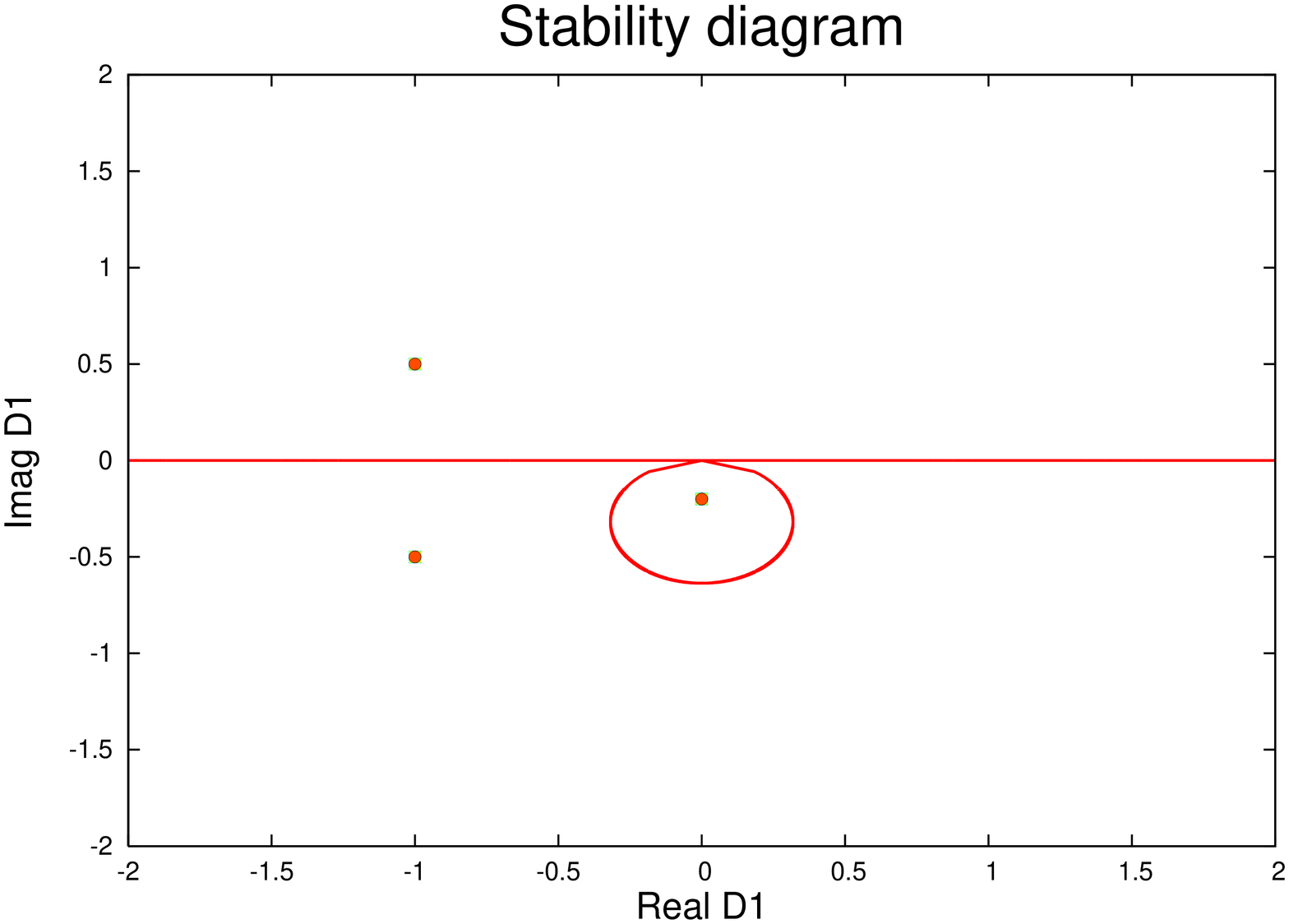}
\includegraphics*[width=60.0mm,height=76.0mm,angle=-90]{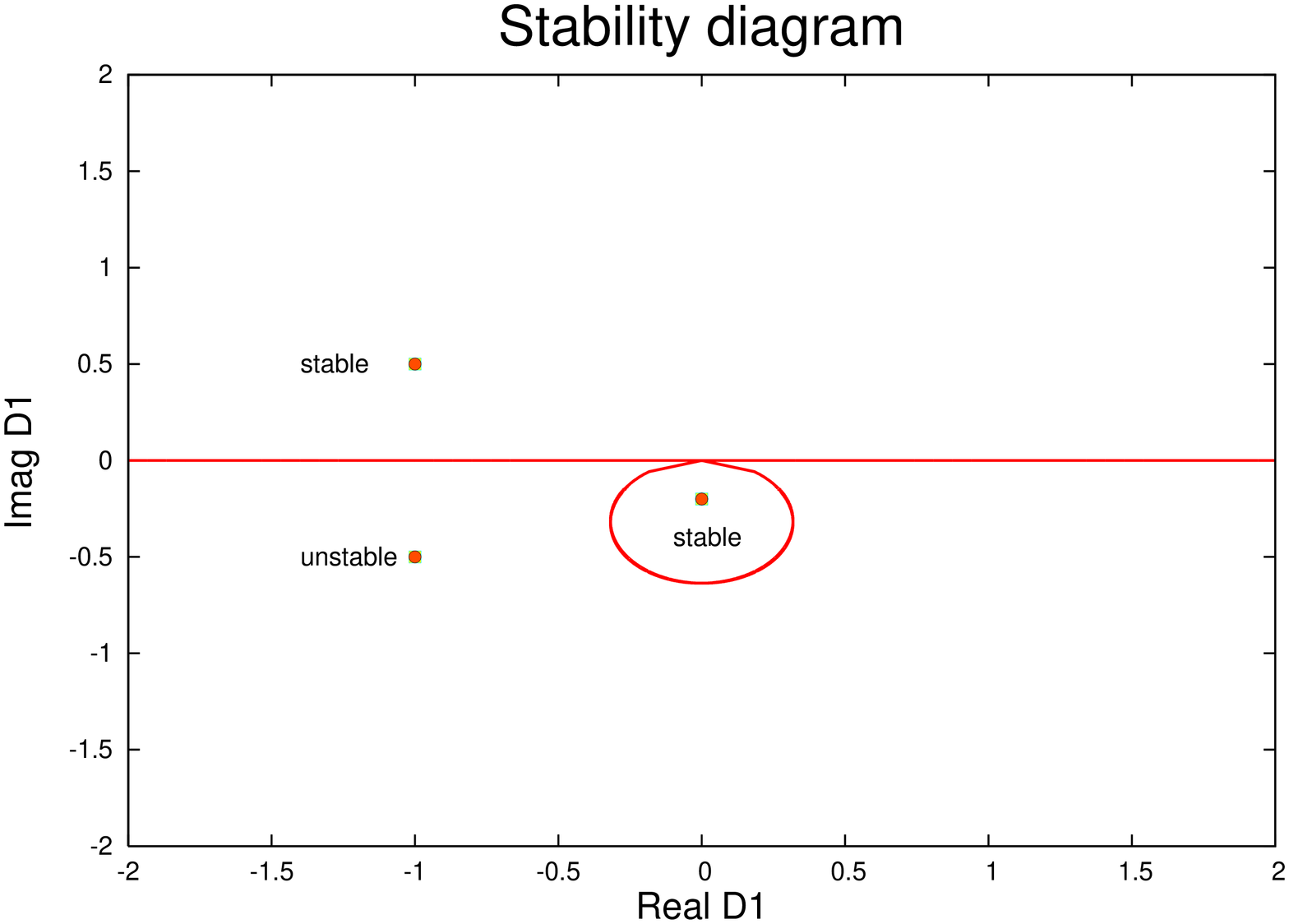}
\end{center}
\caption{Stability diagrams for rectangular distribution together with examples for complex tune shifts (left); stable and unstable points are indicated (right).}
\label{fig:fig14}
\end{figure}
In \Fref{fig:fig14}, we show the same stability diagram together with examples of complex tune shifts.
The stable and unstable points in the stability diagram are indicated in the right-hand side of \Fref{fig:fig14}.

We can use other types of frequency distributions, for example
a bi-Lorentz distribution $\rho(\omega)$.
\begin{figure}[t]
\begin{center}
\includegraphics*[width=60.0mm,height=76.0mm,angle=-90]{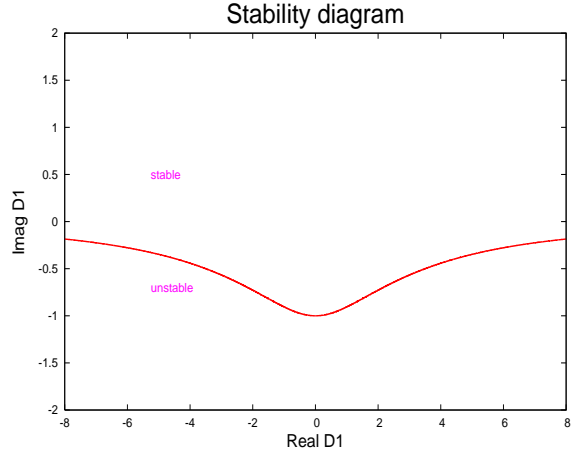}
\end{center}
\caption{Stability diagram for bi-Lorentz frequency distribution. Stable and unstable regions are indicated}
\label{fig:fig15}
\end{figure}
We follow the same procedure as above and the result is shown in \Fref{fig:fig15}.
It can be shown that in all cases half of the complex plane is stable without Landau damping, as indicated in \Fref{fig:fig15}.

\subsection{Examples for unbunched beams}
A similar treatment can be applied to unbunched beams, although some care has to be taken, in
particular in the case of longitudinal stability.

\subsubsection{Transverse unbunched beams}
\begin{figure}[t]
\begin{center}
\includegraphics*[width=70.0mm,height=70.0mm,angle=-90]{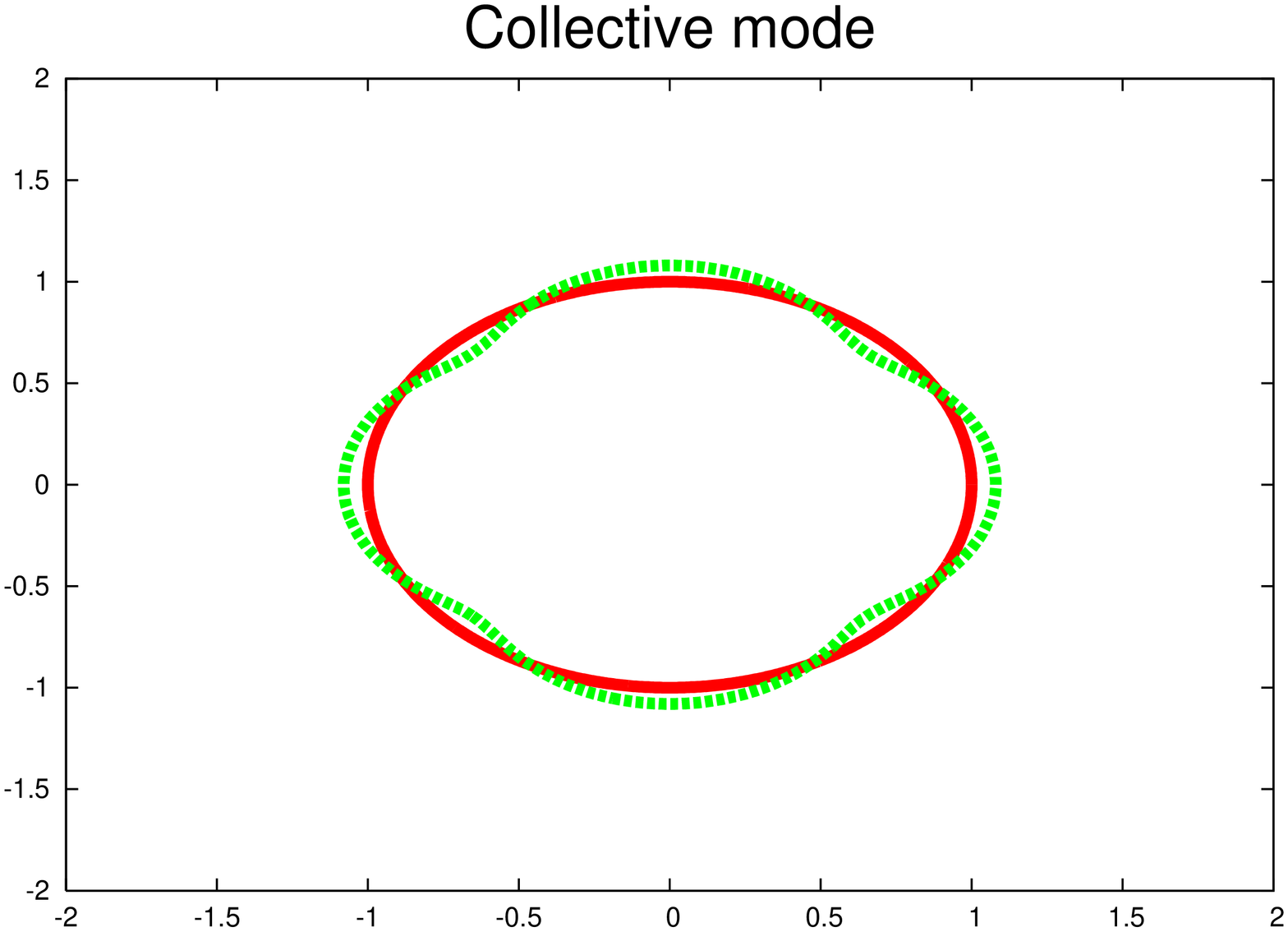}
\includegraphics*[width=70.0mm,height=70.0mm,angle=-90]{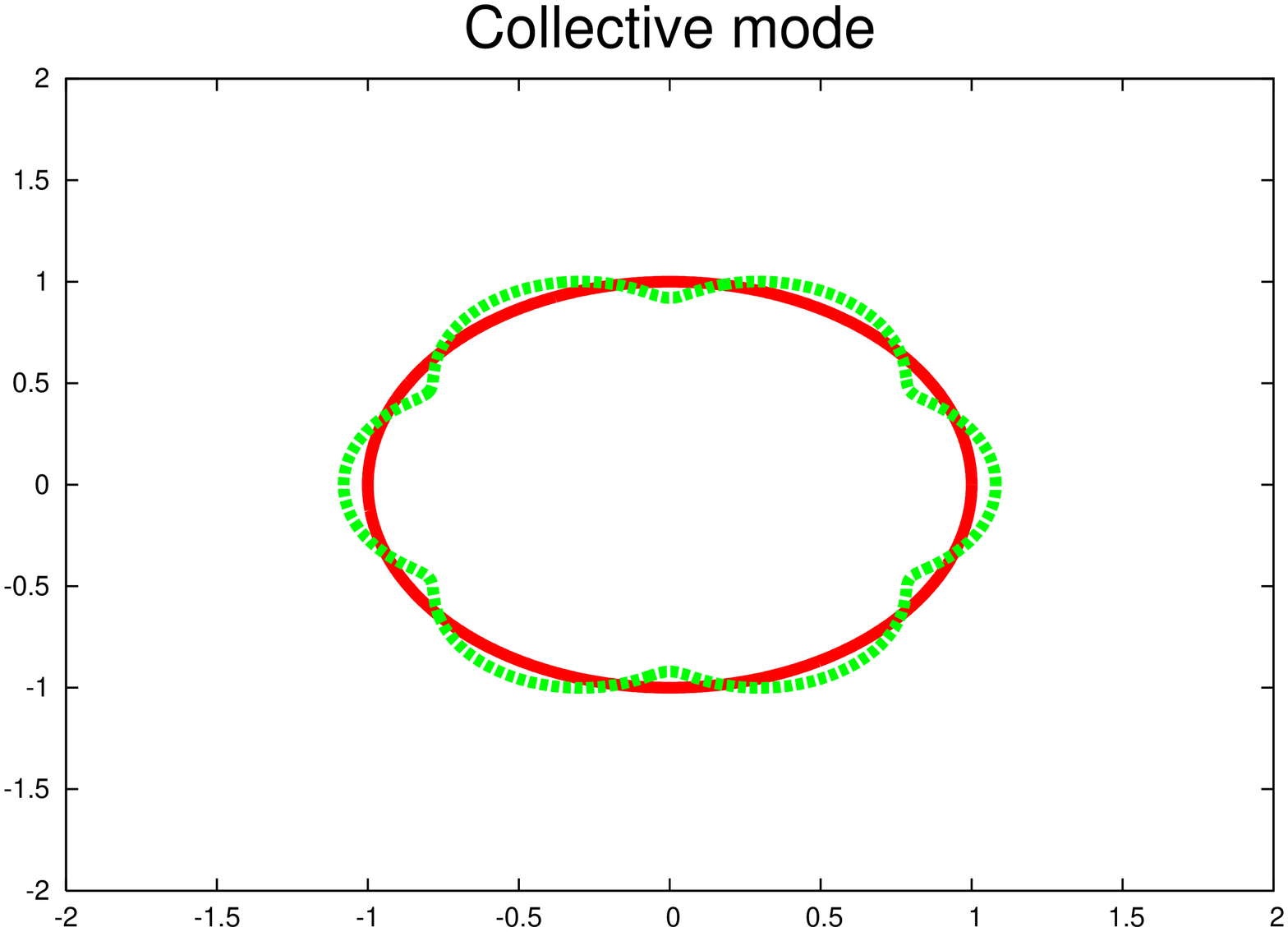}
\end{center}
\caption{Transverse collective mode with mode index $n = 4$ and $n = 6$}
\label{fig:fig16}
\end{figure}

The technique applies directly. Frequency (tune) spread is from:
\begin{itemize}
\item[i)] change of revolution frequency with energy spread (momentum compaction);
\item[ii)] change of betatron frequency with energy spread (chromaticity).
\end{itemize}
The oscillation depends on the mode number $n$ (number of oscillations around the circumference $C$):
\begin{eqnarray}
 \propto \exp(-{\rm i}\Omega t {{+ {\rm i}n(s/C)}})
\end{eqnarray}
and the variable $u$ should be written
\begin{eqnarray}
 u = (\omega_{x} + {{n\cdot \omega_{0}}} - \Omega)/\Delta \omega.
\end{eqnarray}
The rest has the same treatment.
Transverse collective modes in an unbunched beam for mode numbers 4 and 6 are shown
in \Fref{fig:fig16}.

\subsubsection{Longitudinal unbunched beams}
What about longitudinal instability of {{unbunched}} beams?
This is a special case since there is no external focusing, therefore
also no spread $\Delta \omega$ of focusing frequencies.
However, we have a spread in revolution frequency, which is directly related to energy,
and energy excitations directly affect the frequency spread:
\begin{eqnarray}
 \frac{{{\Delta \omega_{\rm rev}}}}{\omega_{0}} = -\frac{\eta}{\beta^{2}}\frac{{{\Delta E}}}{E_{0}}.
\end{eqnarray}
The frequency distribution is described by
\begin{eqnarray}
 \rho(\omega_{\rm rev}) \quad {\mathrm{and}} \quad \Delta \omega_{\rm rev}.
\end{eqnarray}

\begin{figure}[t]
\begin{center}
\includegraphics*[width=50.0mm,height=73.0mm,angle=-90]{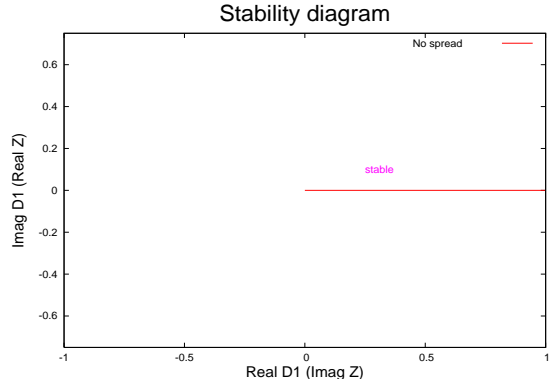}
\end{center}
\caption{Re$(D_{1})$ versus Im$(D_{1})$ for unbunched beam without spread}
\label{fig:fig17}
\end{figure}

Since there is no external focusing ($\omega_{x} = 0$), we have to modify the definition of our parameters:
\begin{eqnarray}
          u = \frac{(\omega_{x} + n\cdot \omega_{0} - \Omega)}{\Delta \omega} \Longrightarrow u = \frac{(n\cdot \omega_{0} - \Omega)}{n\cdot \Delta \omega}
\end{eqnarray}
and introduce two new functions {{$F(u)$}} and {{$G(u)$}}:
the variable $n$ is the mode number.
\begin{eqnarray}
          {{F(u)}} = n\cdot \Delta \omega^{2} {\rm P.V.} \int {\mathrm{d}}\omega_{0} \, \frac{\rho'(\omega_{0})}{n\cdot \omega_{0} - \Omega},
\end{eqnarray}
\begin{eqnarray}
          {{G(u)}} = \pi \Delta \omega^{2} \rho'(\Omega/n)
\end{eqnarray}
to obtain
\begin{eqnarray}
{{-\frac{(\Omega - n\cdot \omega_{0})^{2}}{n^{2} \Delta \omega^{2}}}}  =  {{\frac{1}{(F(u) + {\rm i} G(u))} }}   =  D_{1}.
\end{eqnarray}
As an important consequence, the impedance is now related to the {{square}} of the complex frequency shift $(\Omega - n\cdot \omega_{0})^{2}$.
This has rather severe implications.
\begin{itemize}
\item[i)] Consequence: no more stable region in one half of the plane.
\item[ii)] Landau damping is always required to ensure stability.
\end{itemize}
As an illustration, we show some stability diagrams derived from the new $D_1$.
The stability diagram for unbunched beams, for the longitudinal motion, and without spread, is shown in \Fref{fig:fig17}.
The stable region is just an infinitely narrow line for Im$(D_{1}) = 0$ and positive Re$(D_{1})$.

\begin{figure}[t]
\begin{center}
\includegraphics*[width=50.0mm,height=73.0mm,angle=-90]{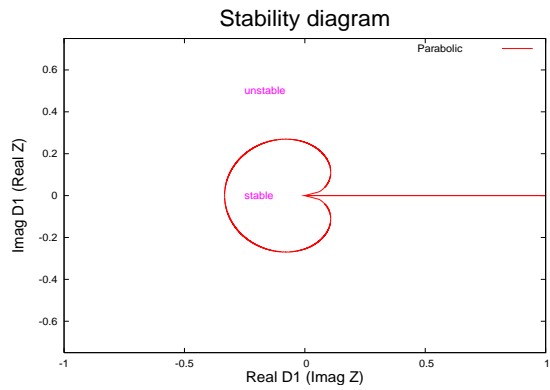}
\end{center}
\caption{Re$(D_{1})$ versus Im$(D_{1})$ for parabolic distribution $\rho(\omega)$ and unbunched beam}
\label{fig:fig18}
\end{figure}

\begin{figure}[t]
\begin{center}
\includegraphics*[width=50.0mm,height=73.0mm,angle=-90]{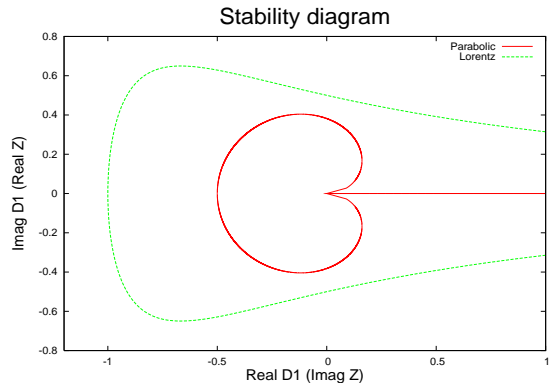}
\end{center}
\caption{Re$(D_{1})$ versus Im$(D_{1})$ for parabolic and Lorentz distributions $\rho(\omega)$ and unbunched beam}
\label{fig:fig19}
\end{figure}

Introducing a frequency spread for a parabolic distribution, we have the stability diagram
shown in \Fref{fig:fig18}.
The locus of the diagram is now the stable region.
As in the previous example, we treat again a Lorentz distribution $\rho(\omega)$ and we show both
in \Fref{fig:fig19}.
The stable region from the Lorentz distribution is significantly larger.
We can investigate this by studying the distributions.
\begin{figure}[t]
\begin{center}
\includegraphics*[width=50.0mm,height=73.0mm,angle=-90]{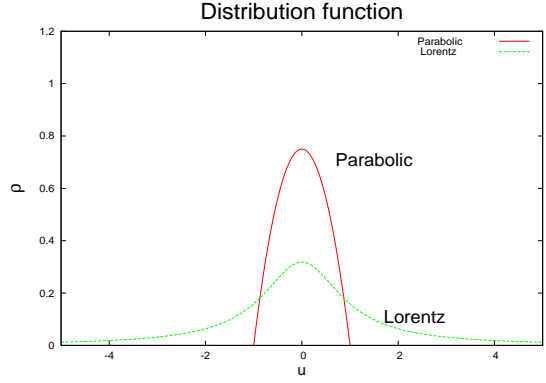}
\end{center}
\caption{Frequency distribution for parabolic and Lorentz distributions $\rho(\omega)$ and unbunched beam}
\label{fig:fig20}
\end{figure}
Both distributions (normalized) are displayed in \Fref{fig:fig20}.
With a little imagination, we can assume that the difference is due to the very
different populations of the tails of the distribution.

A particular problem we encounter is when we do not know the shape of the distribution.
\begin{figure}[t]
\begin{center}
\includegraphics*[width=50.0mm,height=73.0mm,angle=-90]{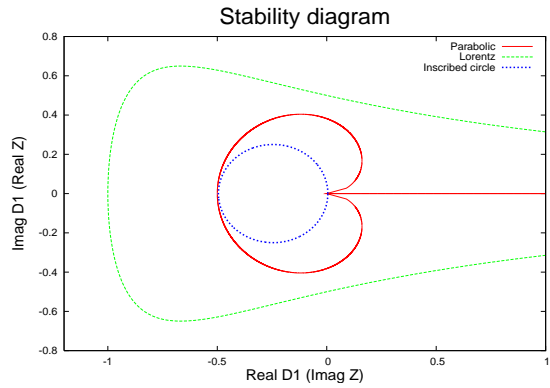}
\end{center}
\caption{Stability diagrams for parabolic and Lorentz distributions, a circle inscribed}
\label{fig:fig21}
\end{figure}
In \Fref{fig:fig21}, we show the stability diagrams again together with a circle inscribed
inside both distributions.
This is a simplified (and pessimistic) criterion for the longitudinal stability of
unbunched beams \cite{bib:keil}.
Since $D_1$ is directly related to the machine impedance $Z$, we obtain a criterion which can be
readily applied for estimating the beam stability.
For longitudinal stability/instability, we have the condition
\begin{eqnarray}
   \frac{|Z_{\parallel}|}{n}  \leq  F \frac{\beta^{2} E_{0} |\eta_{\rm c}|}{q I} \left(\frac{\Delta p}{p}\right)^{2}.
\end{eqnarray}
This is the Keil--Schnell criterion~\cite{bib:keil} applicable to other types of distributions and frequently
used for an estimate of the maximum allowed impedance.
We have the frequency spread from momentum spread and momentum compaction $\eta_{\rm c}$
related to the machine impedance.
For given beam parameters, this defines the maximum allowed impedance $\frac{|Z_{\parallel}|}{n}$.

\section{Effect of simplifications}
We have used a few simplifications in the derivation.
\begin{itemize}
\item[i)] Oscillators are linear.
\item[ii)] Movement of the beam is rigid (i.e.\ beam shape and size do not change).
\end{itemize}
What if we consider the `real' cases,  i.e. non-linear oscillators?
Consider now a bunched beam; because of the synchrotron oscillation,
revolution frequency and betatron spread (from chromaticity) average out.
As sources for the frequency spread we have non-linear forces, such as the following examples.
\begin{itemize}
\item[i)] Longitudinal: sinusoidal RF wave.
\item[ii)] Transverse: octupolar or high multipolar field components.
\end{itemize}
Can we use the same considerations as for an ensemble of {\textit{linear}} oscillators?

The excited betatron oscillation will {{change}} the frequency distribution $\rho(\omega)$ (frequency depends now on the oscillation amplitude).
An oscillating bunch changes the tune (and $\rho(\omega)$) due to the detuning in the non-linear fields.

A complete derivation can be done through the Vlasov equation~\cite{bib:sagan}, but this is well beyond the scope of this article.
The equation
\begin{eqnarray}
 \langle x(t)\rangle  = \frac{A}{2 \omega_{x}} {\rm e}^{-{\rm i}\Omega t} \left[ {\rm P.V.} \int {\mathrm{d}}\omega \,\frac{\rho(\omega)}{(\omega - \Omega)} + {\rm i}\pi \rho(\Omega) \right]
\end{eqnarray}
becomes
\begin{eqnarray}
 \langle x(t)\rangle  = \frac{A}{2 \omega_{x}} {\rm e}^{-{\rm i}\Omega t} \left[ {\rm P.V.} \int {\mathrm{d}}\omega \,\frac{{{\partial\rho(\omega)/\partial \omega}}}{(\omega - \Omega)}
 + {\rm i}\pi {{\partial\rho(\Omega)/\partial \Omega}} \right].
\end{eqnarray}

The distribution function $\rho(\omega)$ has to be replaced by the derivative $\partial\rho(\omega)/\partial \omega$.
We evaluate now this configuration for instabilities in the transverse plane.\\
Since the frequency {{$\omega$}} depends now on the particle's amplitudes {{$J_{x}$}} and {{$J_{y}$}} (see~\cite{bib:herr}), the expression
\begin{eqnarray}
    \omega_{x}(J_{x}, J_{y}) = \frac{\partial H}{\partial J_{x}}
\end{eqnarray}
is the amplitude-dependent betatron tune (similar for $\omega_{y}$).

We then have to write
\begin{eqnarray}
    \rho (\omega) \Longrightarrow \rho(J_{x}, J_{y}).
\end{eqnarray}
Assuming a periodic force in the horizontal $(x)$ plane and using now the tune (normalized frequency) {{$Q =  {\omega}/{\omega_{0}}$}},
\begin{eqnarray}
    F_{x} = A \cdot \exp (-{\rm i} \omega_{0} Q t),
\end{eqnarray}
the dispersion integral can be written as (see also~\cite{bib:rgo} and references therein)
\begin{eqnarray}
 1 = -\Delta Q_{\rm coh} \int_{0}^{\infty}{\mathrm{d}}J_{x} \int_{0}^{\infty}{\mathrm{d}}J_{y} \frac{J_{x}\frac{\partial \rho(J_{x},J_{y})}{\partial J_{x}}}{Q - Q_{x}(J_{x}, J_{y})}.
\end{eqnarray}
Then we proceed as before to get the stability diagram.

If the particle distribution changes (often as a function of time), obviously the frequency distribution $\rho(\omega)$ changes as well.
\begin{itemize}
\item[i)] Examples: higher-order modes; coherent beam--beam modes.
\item[ii)] Treatment requires solving the Vlasov equation (perturbation theory or numerical integration).
\item[iii)] Instead, we use a pragmatic approach: {\textit{unperturbed stability region}} and {\textit{perturbed complex tune shift}}.
\end{itemize}

\section{Landau damping as a cure}
If the boundary of
\begin{eqnarray}
D_{1}  = {{\frac{1}{(f(u) + {\rm i} g(u))} }}
\end{eqnarray}
determines the stability, can we increase the stable region by either of the following methods?
\begin{itemize}
\item[i)] Modifying the frequency distribution {{$Q_{x}(\omega)$}}, i.e. the distribution of the amplitudes {{$Q_{x}(J_{x}, J_{y})$}} (see definition of $f(u)$ and $g(u)$)?
\item[ii)] Introducing tune spread artificially (octupoles, other high-order fields)?
\end{itemize}
The tune dependence of an octupole ({{$k_{3}$}}) can be written as~\cite{bib:herr}
\begin{eqnarray}
 Q_{x}(J_{x}, J_{y}) = Q_{0} + a \cdot {{k_{3}}} \cdot J_{x} + b \cdot {{k_{3}}} \cdot J_{y}.
\end{eqnarray}

Other sources to introduce tune spread are for example:
\begin{itemize}
\item[i)] space charge;
\item[ii)] chromaticity;
\item[iii)] high-order multipole fields;
\item[iv)] beam--beam effects (colliders only).
\end{itemize}

\subsection{Landau damping in the presence of non-linear fields}
The recipe for `generating' Landau damping is:
\begin{itemize}
\item[i)] for a multipole field, compute detuning $Q(J_{x}, J_{y})$;
\item[ii)] given the distribution $\rho(J_{x},J_{y})$;
\item[iii)] compute the stability diagram by scanning frequency, i.e.\ the amplitudes.
\end{itemize}

\subsection{Stabilization with octupoles}
\begin{figure}[t]
\begin{center}
\includegraphics*[width=50.0mm,height=73.0mm,angle=-90]{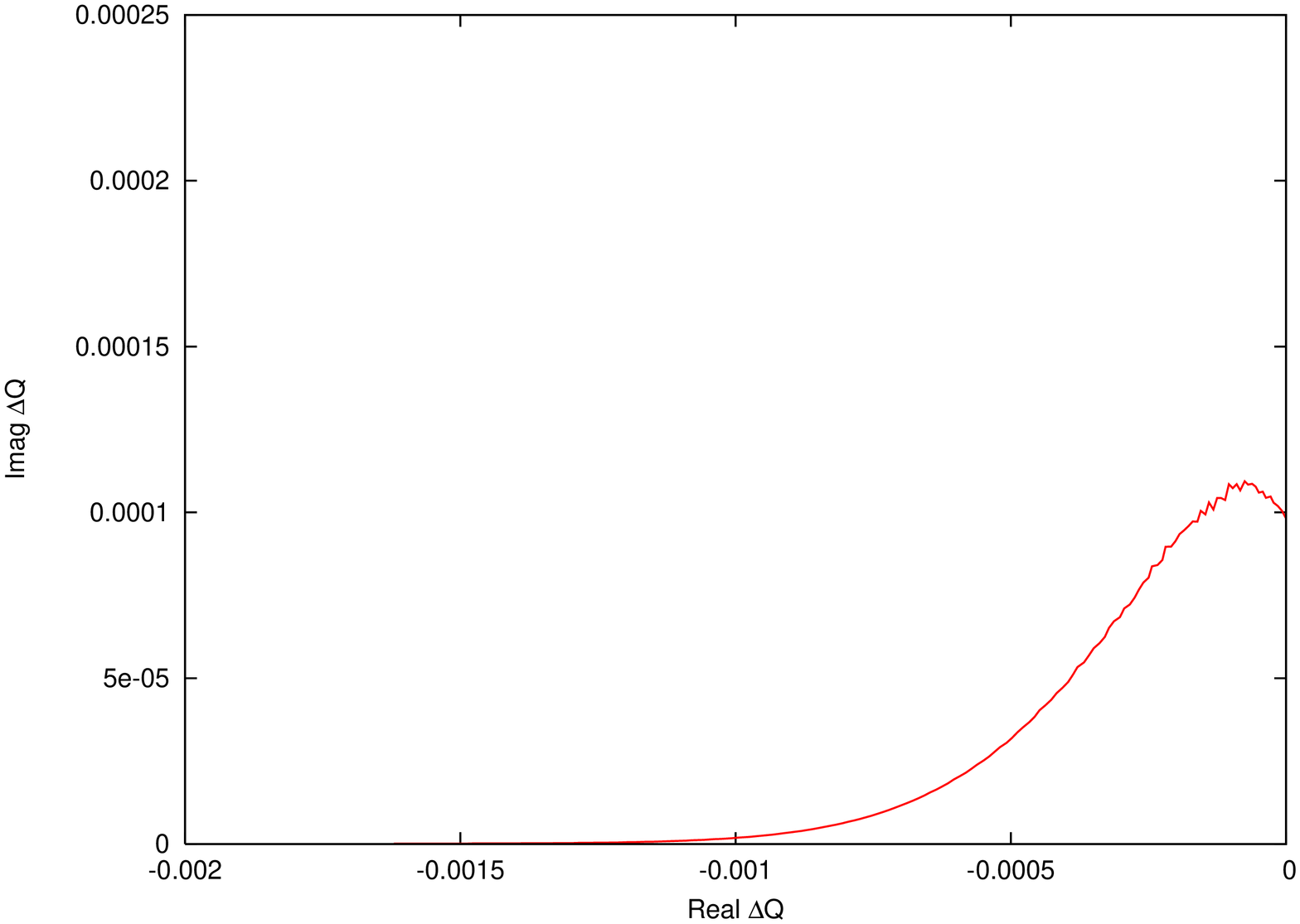}
\includegraphics*[width=50.0mm,height=73.0mm,angle=-90]{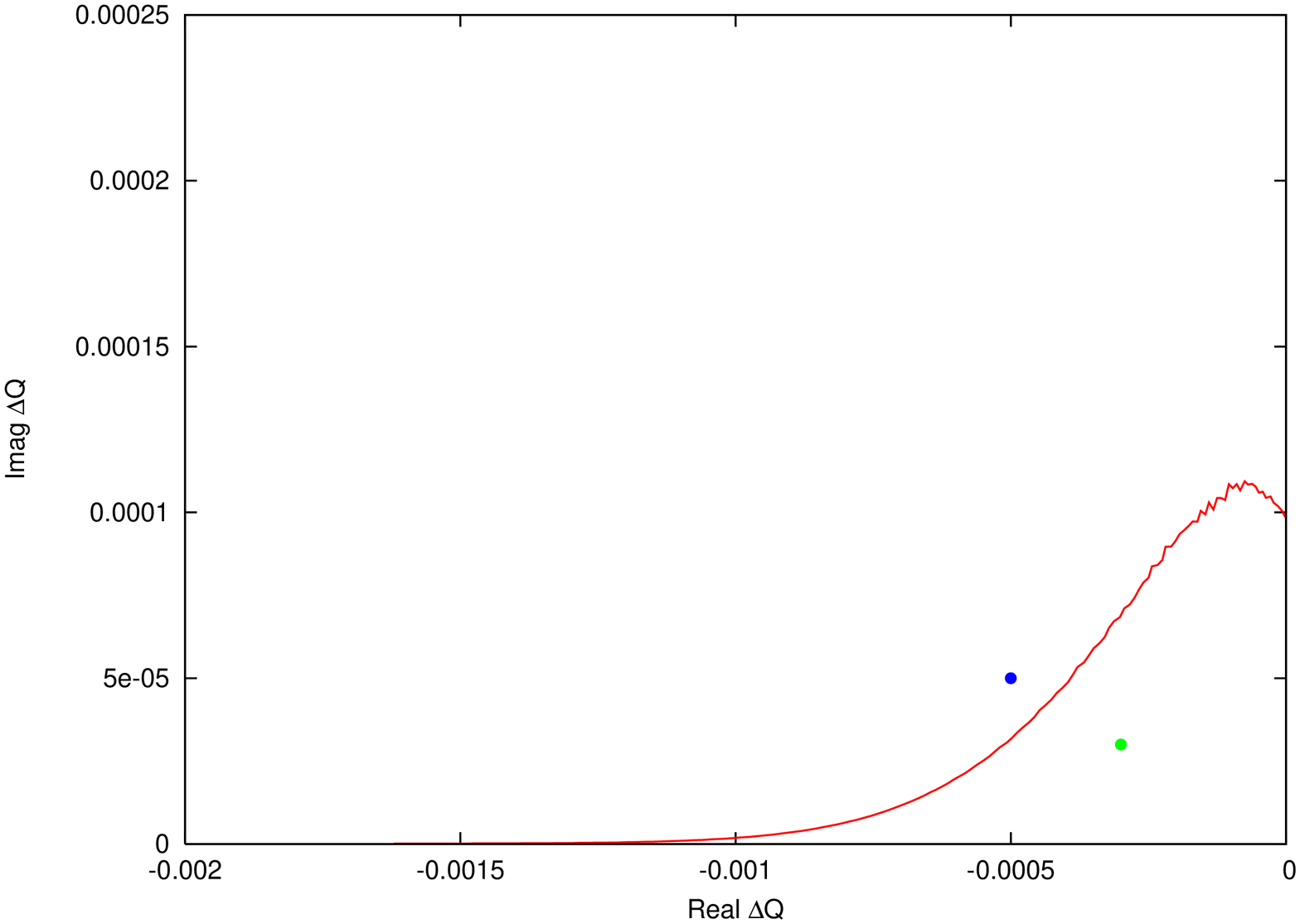}
\end{center}
\caption{Stability diagrams for an octupolar field (left) and together with the complex tune shifts for a stable and an unstable mode (right).}
\label{fig:fig22}
\end{figure}

Figure 21 (left) shows the stability diagram we can get for an octupole.
In Fig. 21 (right), we have added to complex tune shift for a stable and an unstable mode.
The point is inside the region and the beam is stable.
An unstable mode (e.g. after increase of intensity or impedance) is shown in \Fref{fig:fig22} (right); it
lies outside the stable region.

Theoretically, we can increase the octupolar strength to increase the stable region.
However, can we increase the octupole strength as we like?
There are some downsides as follows.
\begin{itemize}
\item[i)] Octupoles introduce strong non-linearities at large amplitudes and may lead to reduced dynamic aperture and bad lifetime.
\item[ii)] We do not have many particles at large amplitudes; this requires large strengths of the octupoles.
\item[iii)] Can cause reduction of dynamic aperture and lifetime.
\item[iv)] They change the chromaticity via feed-down effects!
\item[v)] The lesson: use them if you have no other choice.
\end{itemize}
We see~\cite{bib:herr} that the contribution from octupoles to the stability region
mainly comes from large-amplitude particles, i.e.\ the population of tails.
This has some frightening consequences: changes in the distribution of the tails
can significantly change the stable region.
It is rather doubtful whether tails can be easily reproduced.
Furthermore, in accelerators requiring small losses and long lifetimes (e.g.\ colliders)
the tail particles are unwanted and often caught by the collimation system to protect
the machine.
A reliable calculation of the stability becomes a gamble.
Proposals to artificially increase tails in a hadron collider are a bad choice
although it had been proposed for fast-cycling
machines since they have no need for a long beam lifetime.

\subsection{Example: head--tail modes}
We now apply the scheme to the explicit example of head--tail modes.
They can be due to short-range wake fields or broadband impedance, etc.

The growth and damping times of head--tail modes are usually controlled with {{chromaticity $Q'$}}:
\begin{itemize}
\item[i)] some modes need positive $Q'$;
\item[ii)] some modes need negative $Q'$;
\item[iii)] some modes can be damped by feedback ($m = 0$).
\end{itemize}
In Figs. 22--24,
the stability diagrams are shown for different head--tail modes and different values of the
chromaticity $Q'$.
\begin{figure}[t]
\begin{center}
\includegraphics*[width=50.0mm,height=73.0mm,angle=-90]{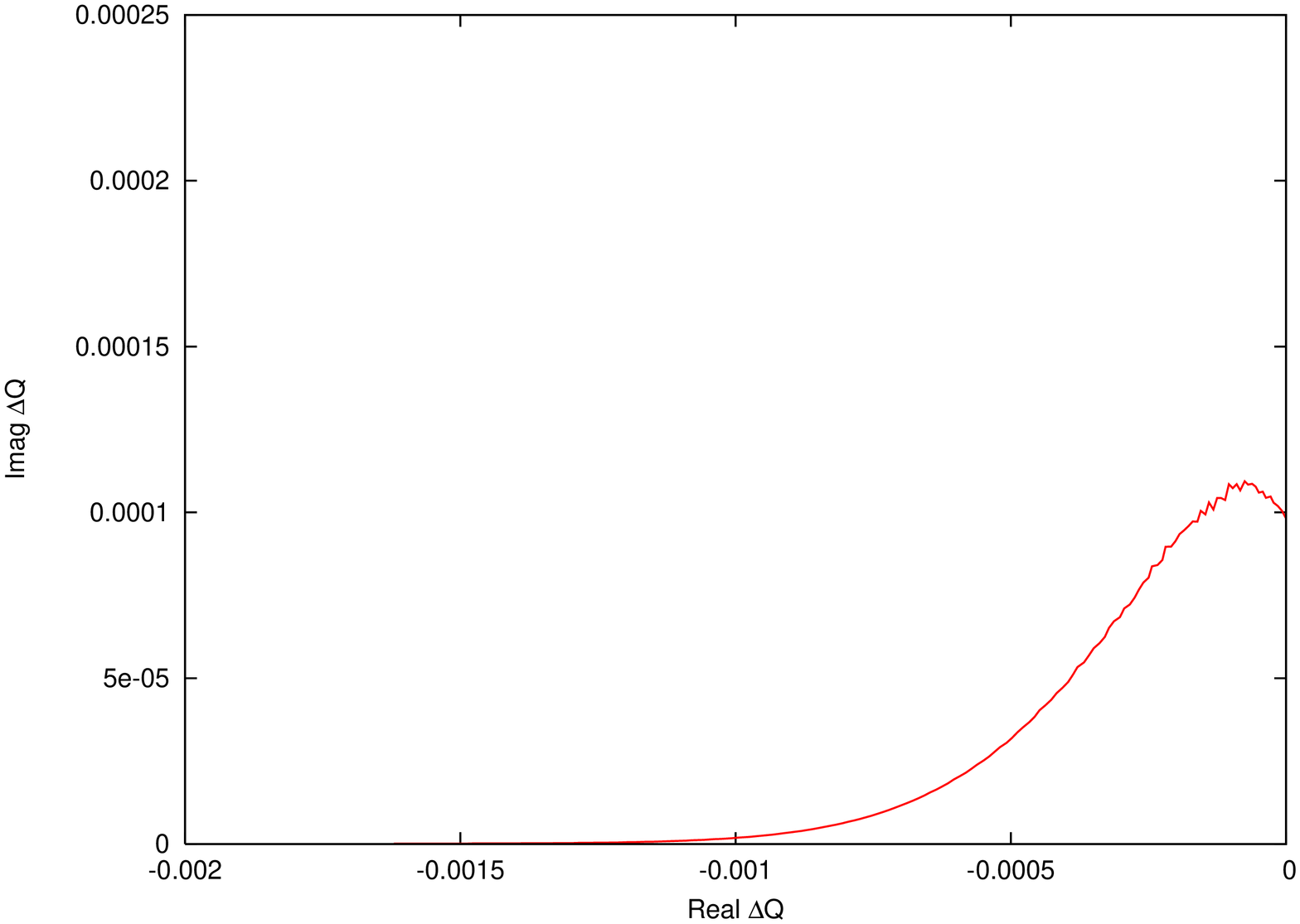}
\includegraphics*[width=50.0mm,height=73.0mm,angle=-90]{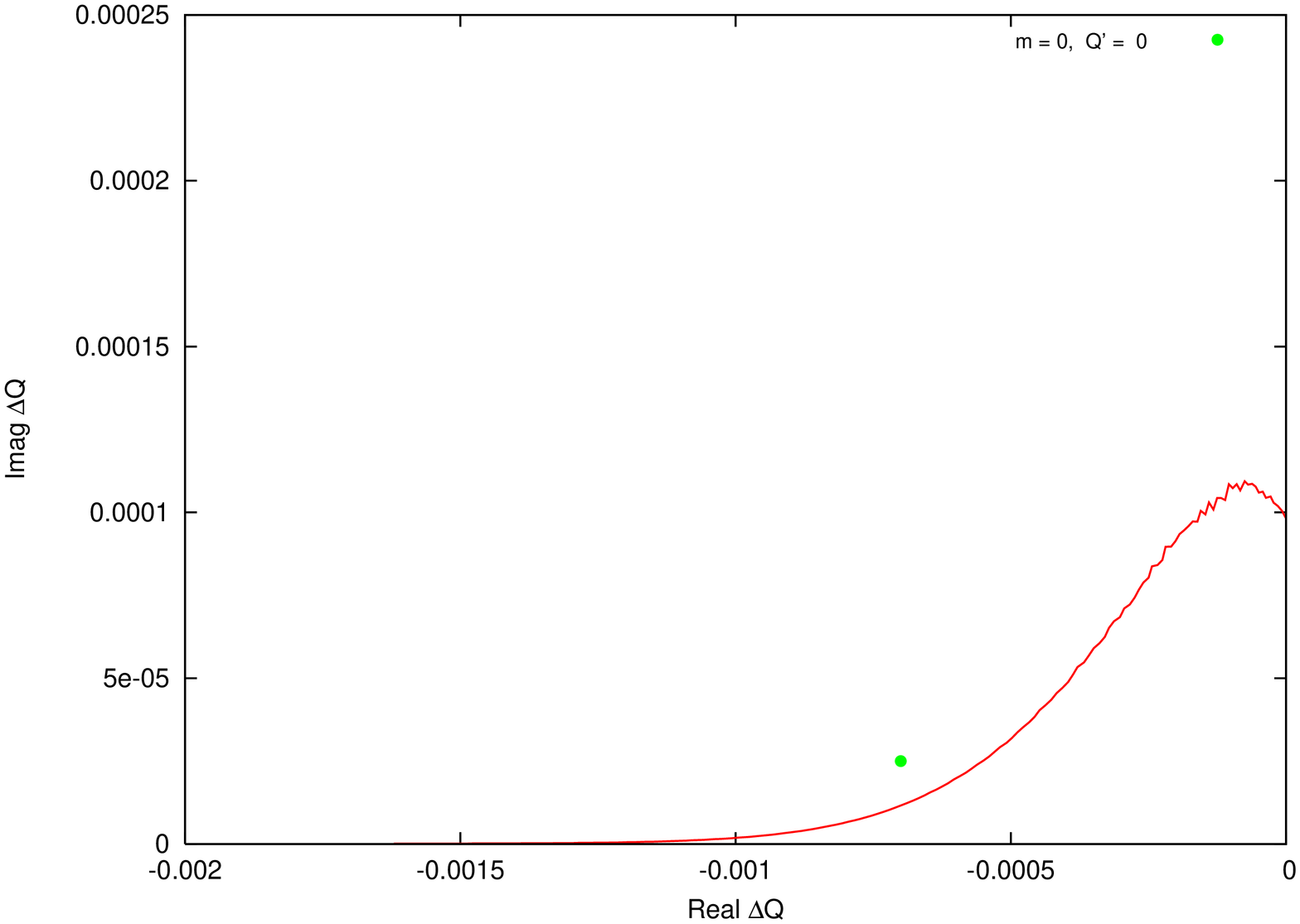}
\end{center}
\caption{Stability diagrams for an octupolar field and $m = 0$ head--tail mode with $Q' = 0$}
\label{fig:fig25}
\end{figure}
For a zero chromaticity $Q' = 0$, the $m=0$ mode is unstable (\Fref{fig:fig25}).
\begin{figure}[t]
\begin{center}
\includegraphics*[width=50.0mm,height=73.0mm,angle=-90]{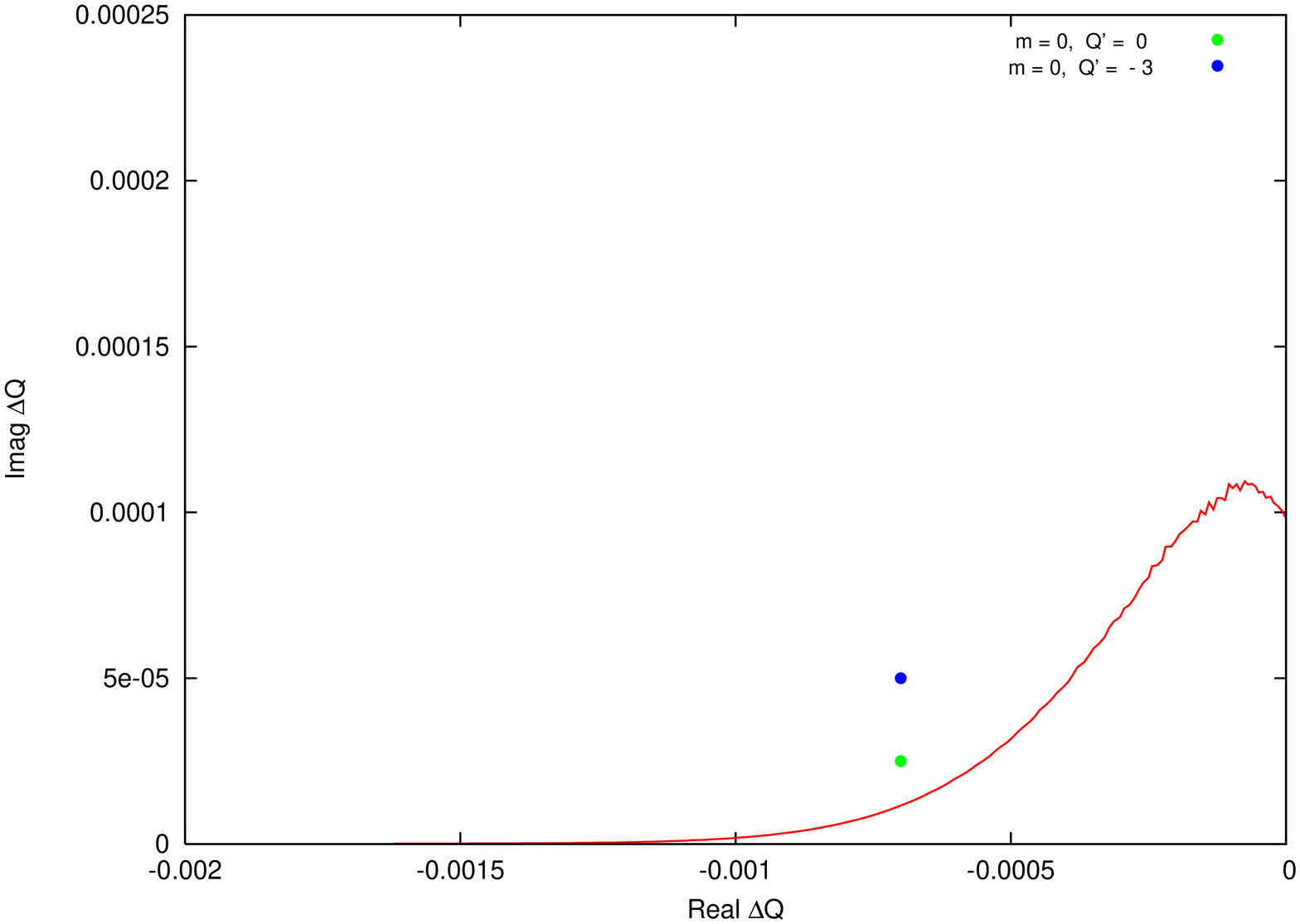}
\includegraphics*[width=50.0mm,height=73.0mm,angle=-90]{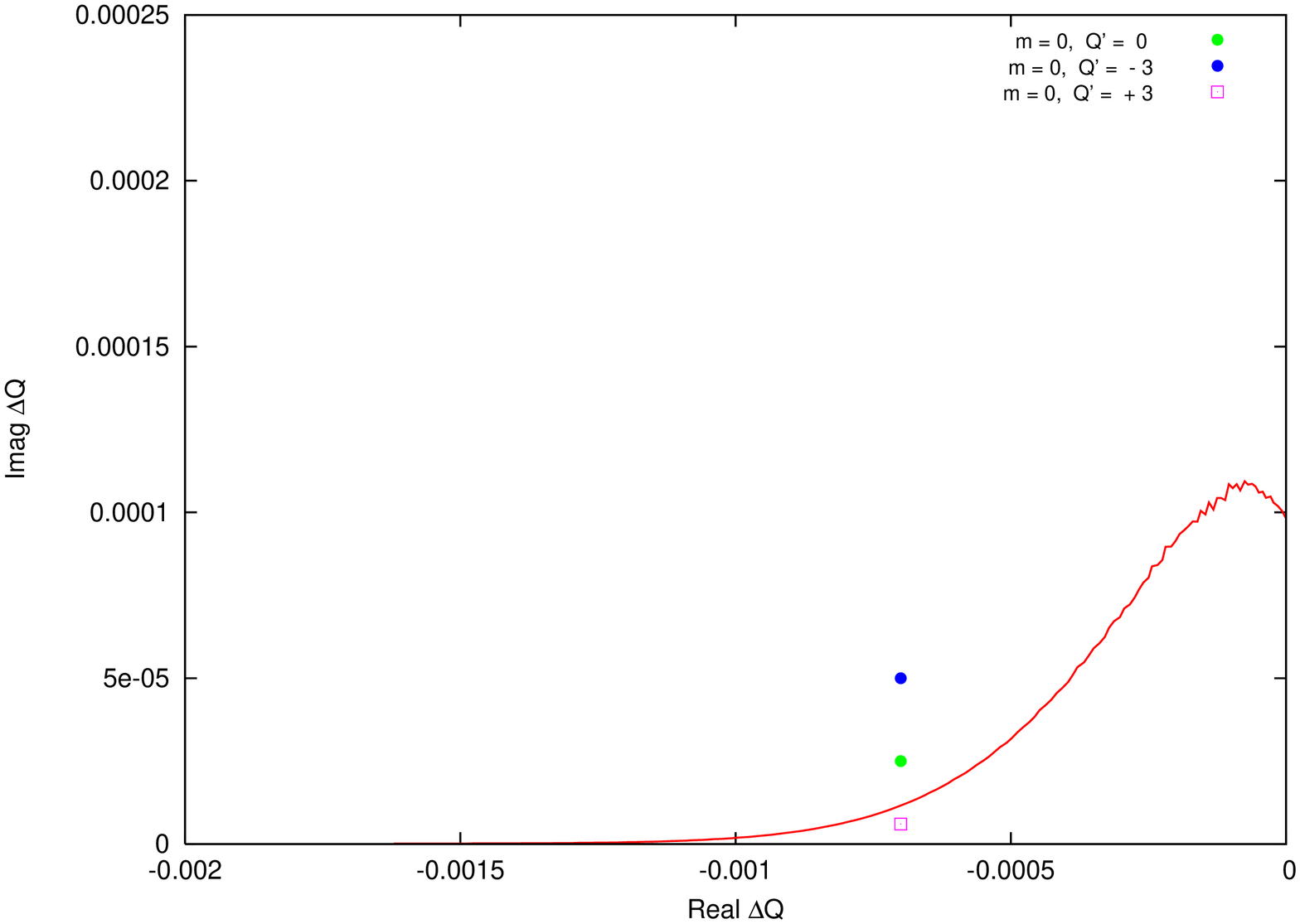}
\end{center}
\caption{Stability diagrams for an octupolar field and $m = 0$ head--tail mode with negative chromaticity $Q' = -3$ (left)
and in the right-hand figure as well a head--tail mode with positive chromaticity $Q' = +3$.}
\label{fig:fig27}
\end{figure}
The positive chromaticity stabilizes the $m=0$ mode by shifting it into the stable area~\Fref{fig:fig27}.
A negative chromaticity makes the beam more unstable by shifting the tune shift further into the unstable region.
However, too large positive chromaticity moves higher order head--tail modes to larger imaginary values, until
they may become unstable (\Fref{fig:fig29}).
\begin{figure}[t]
\begin{center}
\includegraphics*[width=60.0mm,height=90.0mm,angle=-90]{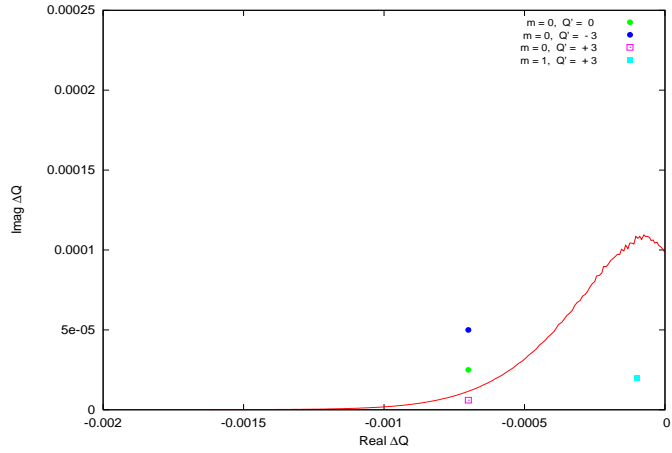}
\end{center}
\caption{Stability diagrams for an octupolar field and $m = 0$ and $m = -1$ head--tail modes with positive chromaticity $Q' = +3$.}
\label{fig:fig29}
\end{figure}
Increasing the chromaticity further than in \Fref{fig:fig29}, the mode $m = -1$ becomes unstable.
Landau damping is required in this case, but with the necessary care to avoid detrimental effects if
this is achieved with octupoles.

\subsection{Stabilization with other non-linear elements}
Can we always stabilize the beam with strong octupole fields?
\begin{itemize}
\item[i)] Would need very large octupole strength for stabilization.
\item[ii)] The known problems:
\begin{itemize}
\item[a)] can cause reduction of dynamic aperture and lifetime;
\item[b)] lifetime important when beam stays in the machine for a long time;
\item[c)] colliders, lifetime more than 10--20 h needed.
\end{itemize}
\item[iii)] Is there another option?
\end{itemize}

\begin{figure}[t]
\begin{center}
\includegraphics*[width=60.0mm,height=90.0mm,angle=-90]{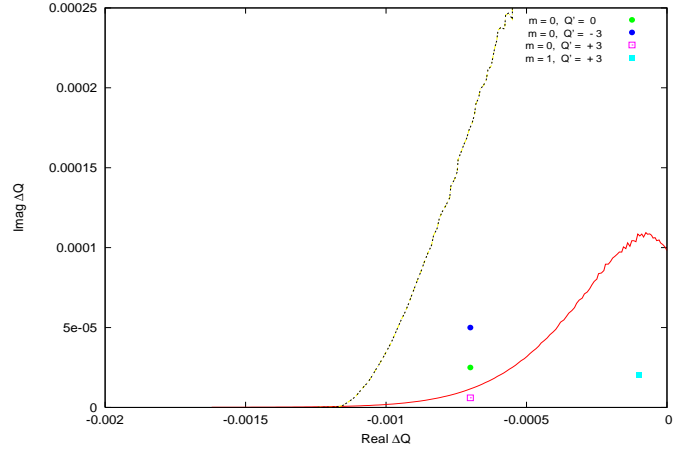}
\end{center}
\caption{Stability diagrams for an octupolar field and head--tail modes (\Fref{fig:fig29}) and stability diagram from beam--beam effects.}
\label{fig:fig30}
\end{figure}

In the case of colliders, the beam--beam effects provide a large tune spread that can
be used for Landau damping~(\Fref{fig:fig30}).
Even when the tune spread is comparable, the stability region from a head-on beam--beam interaction is
much increased~\cite{bib:herrvos}.
We observe a large difference between the two stability diagrams.
Where does this difference come from?
The tune dependence of an octupole can be written as
\begin{eqnarray}
 Q_{x}(J_{x}, J_{y}) = Q_{0} + a J_{x} + b J_{y}
\end{eqnarray}
and is linear in the action $J$ (for coefficients, see Appendix \ref{sec:app2}).

\begin{figure}[t]
\begin{center}
\includegraphics*[width=80.0mm,height=80.0mm,angle=+00]{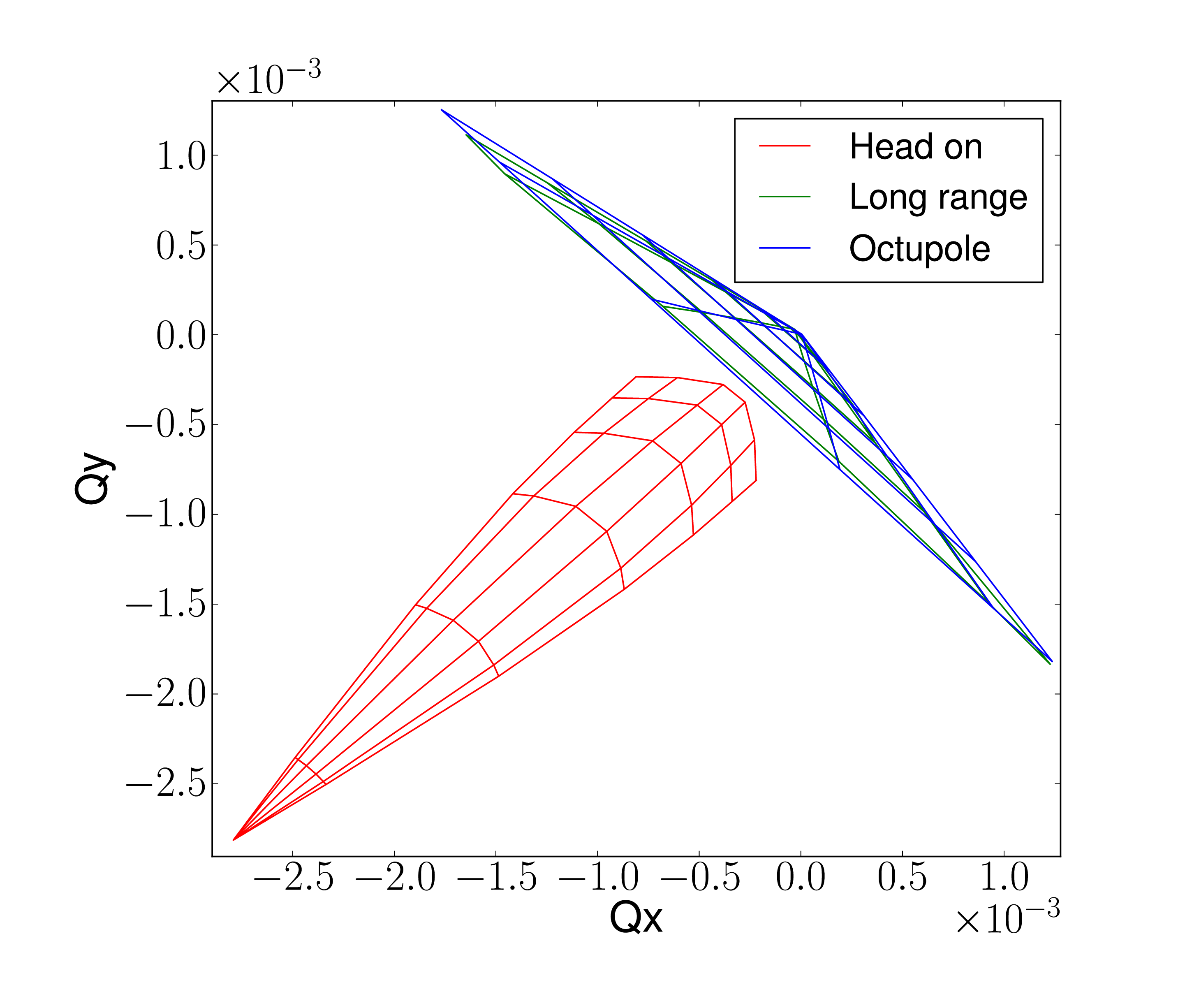}
\end{center}
\caption{Tune spread in two dimensions for octupoles, long-range beam--beam and head-on beam--beam. Tune spread adjusted to be the same for all cases.}
\label{fig:fig31}
\end{figure}

The tune dependence of a head-on beam--beam collision can be written~\cite{bib:herrbb, bib:pieloni} as follows:
with ${{\alpha}} = {x}/{\sigma^{*}}$, we get $\Delta Q/\xi = ({4}/{{{\alpha^{2}}}})\left[ 1 - I_{0}({{{\alpha^{2}}}}/{4})\cdot {\rm e}^{{-{{\alpha^{2}}}}/{4}}\right]$.
As a demonstration, we show in \Fref{fig:fig31} the tune spread from octupoles,
long-range beam--beam effects~\cite{bib:herrbb} and head-on beam--beam effects.
The parameters are adjusted such that the overall tune spread is always the same for the three cases.
However, the head-on beam--beam tune spread has the largest effect for {\textit{small}} amplitudes,
i.e.\ where the particle density is largest and the contribution to the stability is strong.
\begin{figure}[t]
\begin{center}
\includegraphics*[width=85.0mm,height=70.0mm,angle=-00]{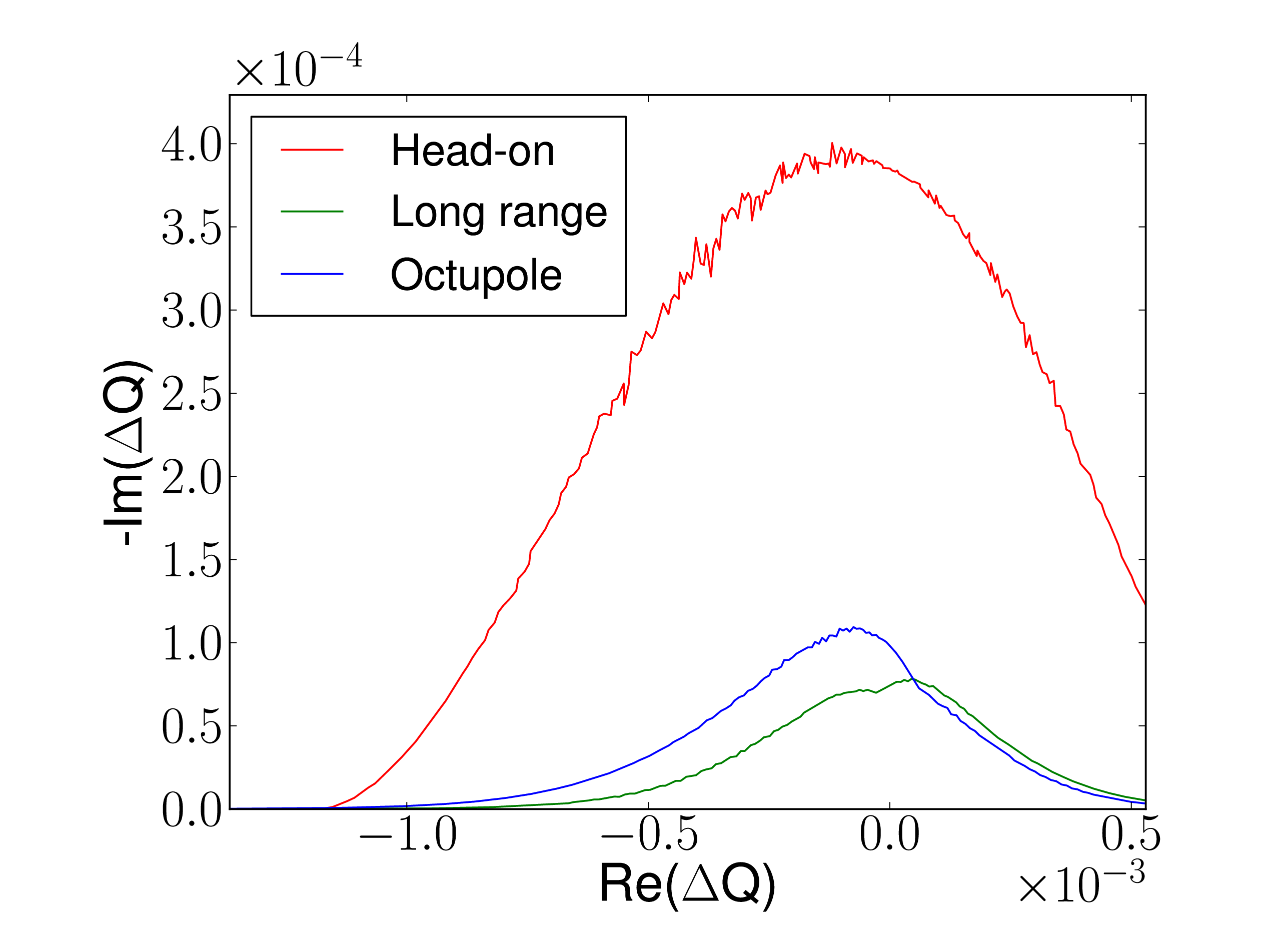}
\end{center}
\caption{Stability diagrams for octupoles, long-range beam--beam and head-on beam--beam. Tune spread adjusted to be the same for all cases.}
\label{fig:fig32}
\end{figure}
For the tune spreads shown in \Fref{fig:fig31}, we have computed the corresponding stability diagrams~\cite{bib:buffat}.
Although the tune spread is the same, the stability region shown in \Fref{fig:fig32} is significantly larger.
As another and most important advantage, since the stability is ensured by small-amplitude particles,
the stable region is very insensitive to the presence of tails or the exact distribution function.
Special cases such as collisions of `hollow beams' are not noteworthy in this overview.

\subsubsection{Conditions for Landau damping}
A tune spread is not sufficient to ensure Landau damping.
To summarize, we need:
\begin{itemize}
\item[i)] presence of incoherent frequency (tune) spread;
\item[ii)] coherent mode must be {{inside}} this spread;
\item[iii)] the same particles must be involved in the oscillation and the spread.
\end{itemize}
The last item requires some explanation.
When the tune spread is not provided by the particles actually involved in the coherent motion, it is
not sufficient and results in no damping.
For example, this is the case for coherent beam--beam modes where the eigenmodes of coherent beam--beam modes driven either
by head-on effects or long range interactions are very different~\cite{bib:alexahin1, bib:alexahin2, bib:alexahin3}.
For modes driven by the head-on interactions, mainly particles at small amplitudes are participating in
the oscillation. The tune spread produced by long range interactions produces little damping in this case.

\begin{figure}[t]
\begin{center}
\includegraphics*[width=50.0mm,height=50.0mm,angle=-00]{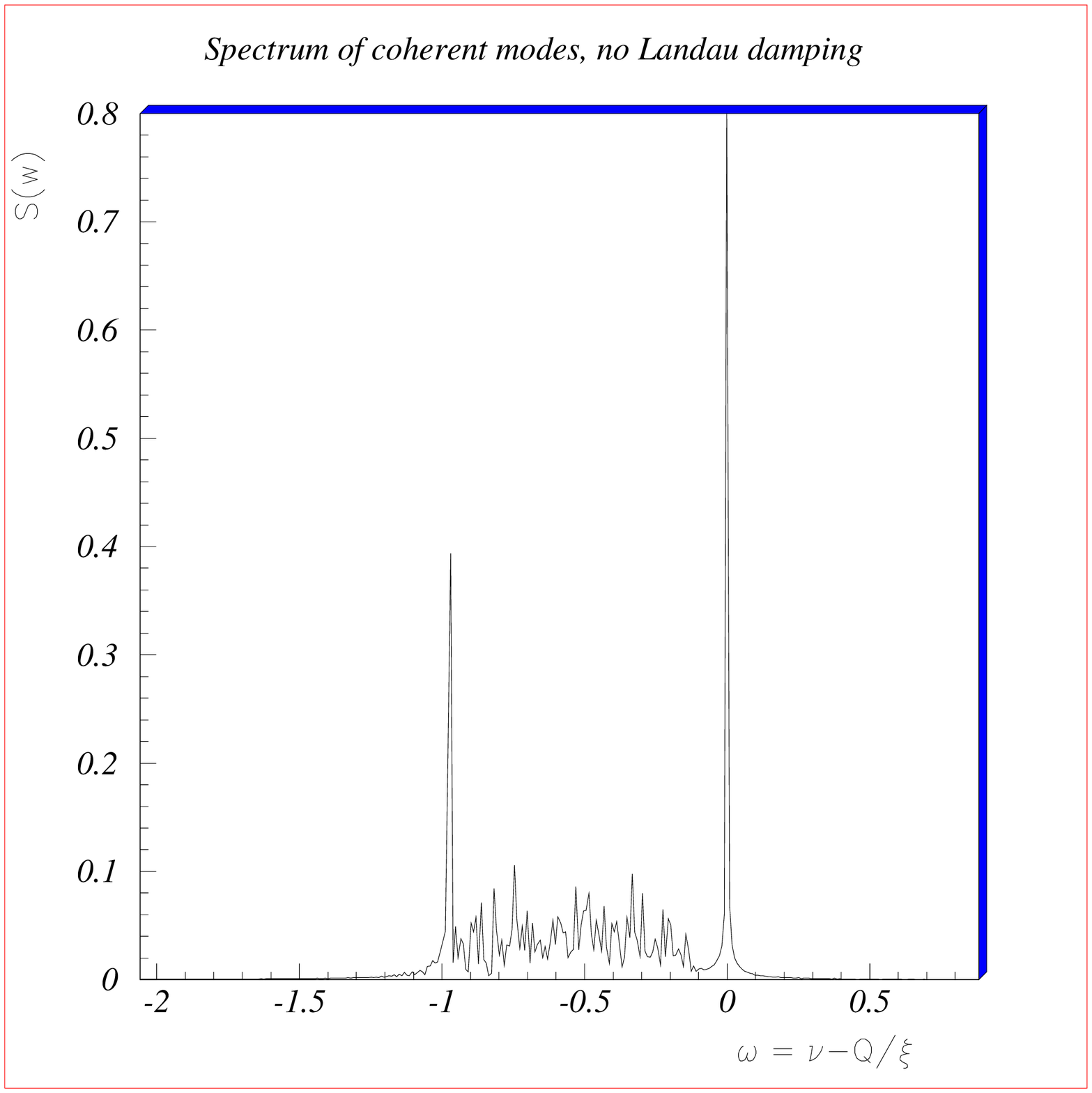}
\includegraphics*[width=50.0mm,height=50.0mm,angle=-00]{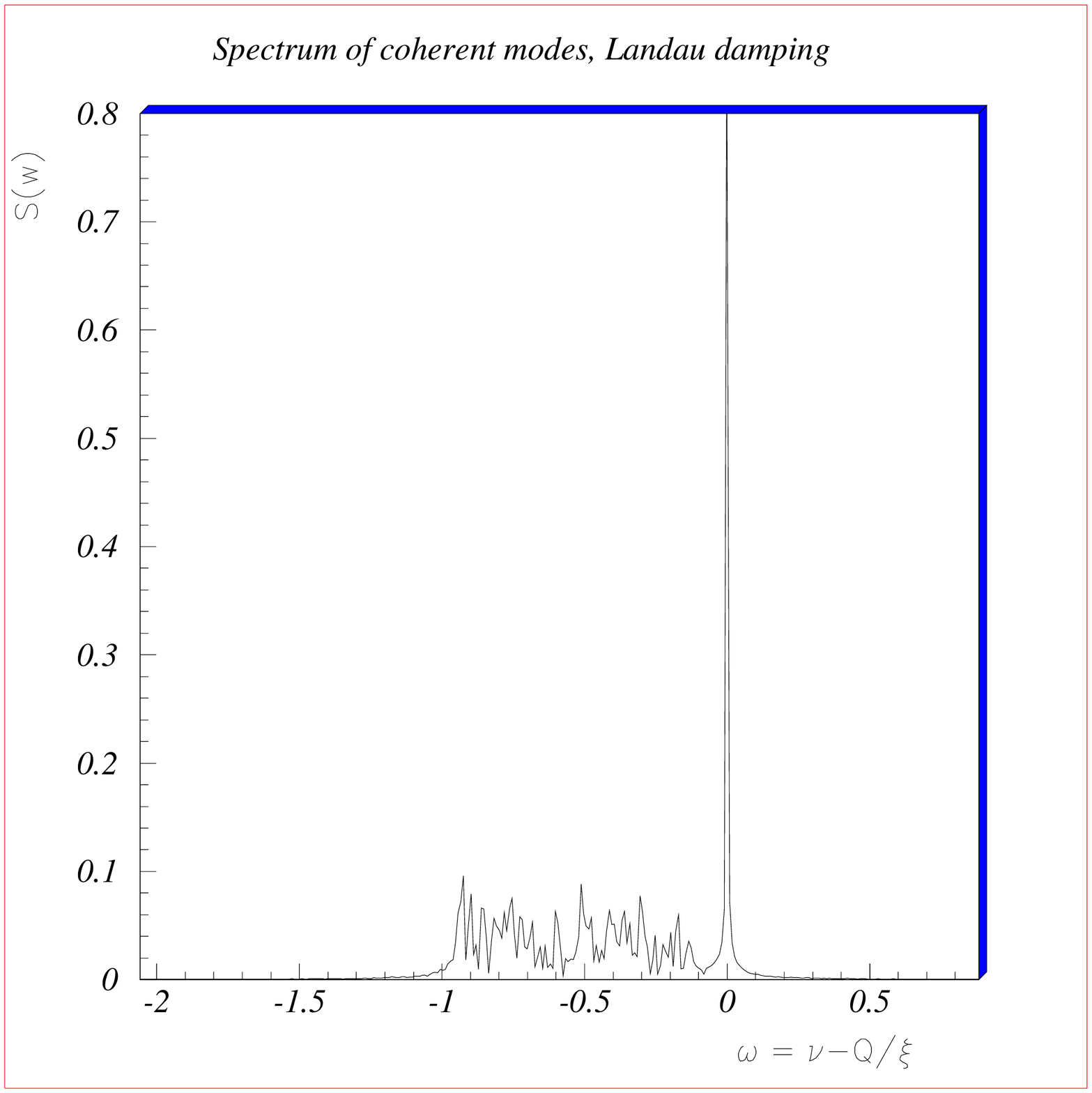}
\end{center}
\caption{Coherent beam--beam modes with $\pi$-mode inside and outside the incoherent tune spread due to beam--beam effects.
Decoupling is due to intensity difference between the two beams.}
\label{fig:fig33}
\end{figure}

Another option to damp coherent beam--beam modes is shown in \Fref{fig:fig33}.
It shows the results of a simulation for the main coherent beam--beam modes (0-mode and $\pi$-mode)
for slightly different intensities.
The difference is sufficient to decouple the two beams and they cannot maintain the correct phase.
We have:
\begin{itemize}
\item[i)]coherent mode inside or outside the incoherent spectrum, depending on the intensity difference;
\item[ii)]Landau damping restored when the symmetry is sufficiently broken.
\end{itemize}

\subsection{Landau damping with non-linear fields: are there any side effects?}
Introducing non-linear fields into an accelerator is not always a desirable
procedure.
It may have implications for the beam dynamics and we separate them into three
categories.
\begin{itemize}
\item[i)] Good:
\begin{itemize}
\item[a)] stability region increased.
\end{itemize}
\item[ii)] Bad:
\begin{itemize}
\item[a)] non-linear fields introduced (resonances);
\item[b)] changes optical properties, e.g.\ chromaticity (feed-down).
\end{itemize}
\item[iii)] Special cases:
\begin{itemize}
\item[a)] non-linear effects for large amplitudes (octupoles);
\item[b)] much better: head-on beam--beam (but only in colliders).
\end{itemize}
\end{itemize}
Landau damping with non-linear fields is a very powerful tool, but the side effects and
implications have to be taken into account.

\section{Summary}
The collisionless damping of coherent oscillations as predicted by Landau in 1946 is
an important concept in plasma physics as well as in other applications such as
hydrodynamics, astrophysics and biophysics to mention a few.
It is used extensively in particle accelerators to avoid coherent oscillations of
the beams and the instabilities.
Despite its intensive use, it is not a simple phenomenon and the interpretation
of the physics behind the mathematical structures is a challenge.
Even after many decades after its discovery, there is (increasing) interest in this
phenomenon and work on the theory (and extensions of the theory) continues.

\appendix
\section{Solving the dispersion relation}
\label{sec:app1}
\subsection{Dispersion relation from Vlasov's calculation}
Starting with the dispersion relation derived using Vlasov's approach~(\ref{eq:04}):
\begin{eqnarray}
     1 + \frac{\omega_{\rm p}^{2}}{k^{2}} \int \frac{\partial \psi_{0}/\partial v}{(\frac{\omega}{k} - v)}\, {\mathrm{d}}v = 0,
\end{eqnarray}
we have to make a few assumptions.
We assume that we can restrict ourselves to waves with $\omega/k \gg v$ or $\omega/k \ll v$.
The latter case cannot occur in Langmuir waves and we can assume the case with $\omega/k \gg v$.
Then we can integrate the integral by parts and get
\begin{eqnarray}
     1 + \frac{\omega_{\rm p}^{2}}{k^{2}} \int \frac{\psi_{0}}{( {\omega}/{k} - v)^{2}}\, {\mathrm{d}}v = 0
\end{eqnarray}
or, rewritten for the next step,
\begin{eqnarray}
     1 + \frac{\omega_{\rm p}^{2}}{\omega^{2}} \int \frac{\psi_{0}}{(1 - {v k}/{\omega})^{2}}\, {\mathrm{d}}v = 0.
\end{eqnarray}
With the assumption $\omega/k \gg v$, we can expand the denominator in series of $\frac{v k}{\omega}$ and obtain
\begin{eqnarray}
     1 + \frac{\omega_{\rm p}^{2}}{\omega^{2}} \int \psi_{0} {\mathrm{d}}v \, \cdot \left(1 + 2\cdot\left(\frac{v k}{\omega}\right) + 3\cdot\left(\frac{v k}{\omega}\right)^{2} \right)= 0.
\end{eqnarray}
For the next step as an explicit example we use a Maxwellian velocity distribution, i.e.
\begin{eqnarray}
     \psi(v) = {\frac{1}{\sqrt{2\pi}}} \frac{1}{v_{\rm p}}  {\rm e}^{- {v^{2}}/{2 v_{\rm p}^{2}}}.
\end{eqnarray}
The individual integrals are then
\begin{eqnarray}
     \int \psi(v) \, {\mathrm{d}}v = 1, \quad \int \psi(v)\cdot v \, {\mathrm{d}}v = 0, \quad \int \psi(v)\cdot v^{2} \, {\mathrm{d}}v = v_{\rm p}^{2} = \frac{\omega_{\rm p}^{2}}{k^{2}}
\end{eqnarray}
to obtain finally
\begin{eqnarray}
     1-\frac{\omega_{\rm p}^{2}}{\omega^{2}}-\frac{3k^{2}v_{\rm p}^{2}\omega_{\rm p}^{2}}{\omega^{4}} = 0.
\end{eqnarray}
This dispersion relation can now be solved for $\omega$ and we get two solutions:
\begin{eqnarray}
     \omega^{2} = \frac{1}{2}\omega_{\rm p}^{2}~\pm~\frac{1}{2}\omega_{\rm p}^{2}\left( 1 + \frac{12k^{2}v_{\rm p}^{2}}{\omega_{\rm p}^{2}}\right)^{1/2}.
\end{eqnarray}

Rewritten, we obtain the well-known dispersion relation for Langmuir waves:
\begin{eqnarray}
     \omega^{2} = \omega_{\rm p}^{2}\left( 1 + 3 k^{2} \lambda^{2}\right) \quad (\lambda = v_{\rm p}/\omega_{\rm p}).
\label{eq:b01}
\end{eqnarray}
The frequency is real and there is no damping.

\subsection{Dispersion relation from Landau's approach}
Here we solve the dispersion relation using the result obtained by Landau~(\ref{eq:05}):
\begin{eqnarray}
     1 + \frac{\omega_{\rm p}^{2}}{k} \left[ {\rm P.V.}\int \frac{\partial \psi_{0}/\partial v}{(\omega - k v)}\, {\mathrm{d}}v
     {- \frac{{\rm i}\pi}{k}\left( \frac{\partial \psi_{0}}{\partial v}\right)_{v = \omega/k}}\right] = 0.
\label{eq:a01}
\end{eqnarray}
This should lead to a damping.

Integration by parts leads now to
\begin{eqnarray}
     1-\frac{\omega_{\rm p}^{2}}{\omega^{2}}-\frac{3k^{2}v_{\rm p}^{2}\omega_{\rm p}^{2}}{\omega^{4}}
     -\frac{{\rm i}\pi}{k}\left( \frac{\partial \psi_{0}}{\partial v}\right)_{v = \omega/k} = 0.
\end{eqnarray}
For the real part, we can use the same reasoning as for Vlasov's calculation and find again
\begin{eqnarray}
     1-\frac{\omega_{\rm p}^{2}}{\omega_{\rm r}^{2}}-\frac{3k^{2}v_{\rm p}^{2}\omega_{\rm p}^{2}}{\omega_{\rm r}^{4}} = 0.
\end{eqnarray}
Now we have used $\omega_{\rm r}$ to indicate that we computed the real part of the complex frequency
\begin{eqnarray}
     \omega = \omega_{\rm r} + {\rm i}\cdot \omega_{\rm i}.
\end{eqnarray}
We assume `weak damping', i.e.\ $\omega_{\rm i} \ll \omega_{\rm r}$.
This leads us to $\omega^{2} \simeq \omega_{\rm r} + 2 {\rm i} \omega_{\rm r} \omega_{\rm i}$.
With the solution (\ref{eq:b01}) and $k \lambda \ll 1$, we find for the imaginary part $\omega_{\rm i}$
\begin{eqnarray}
     \omega_{\rm i} = \frac{\pi\cdot \omega_{\rm p}^{3}}{2\cdot k^{2}}\cdot \left( \frac{\partial \psi_{0}}{\partial v}\right)_{v = \omega/k}.
\end{eqnarray}
Using the Maxwell velocity distribution as an example:
\begin{eqnarray}
     \psi(v) = {\frac{1}{\sqrt{2\pi}}} \frac{1}{v_{\rm p}}  {\rm e}^{-{v^{2}}/{2 v_{\rm p}^{2}}},
\end{eqnarray}
we get for the derivative
\begin{eqnarray}
     \frac{\partial\psi(v)}{\partial v} = -{\frac{1}{\sqrt{2\pi}}} \frac{1}{v_{\rm p}^{3}}  {\rm e}^{- {v^{2}}/{2 v_{\rm p}^{2}}}.
\end{eqnarray}
Using again $\lambda = v_{\rm p}/\omega_{\rm p}$ to simplify the expression, we can expand in a series (because $\omega/k \gg v$) and arrive at
\begin{eqnarray}
     \omega_{\rm i} = -\omega_{\rm p}\cdot \frac{1}{k^{3}\lambda^{3}}\sqrt{\frac{\pi}{2}}\cdot \exp\left(-\frac{1}{2 k^{2} \lambda^{2}} - \frac{3}{2} \right).
\end{eqnarray}
This is the damping term we obtain using Landau's result for plasma oscillations.

\section{Detuning with octupoles}
\label{sec:app2}
The tune dependence of an octupole can be written as~\cite{bib:herr}
\begin{eqnarray}
 Q_{x}(J_{x}, J_{y}) = Q_{0} + a J_{x} + b J_{y}
\end{eqnarray}
for the coefficients:
\begin{eqnarray}
    \Delta Q_{x} = \left[\frac{3}{8 \pi} \int \beta_{x}^{2}\frac{K_{3}}{B \rho} \, {\mathrm{d}}s \right] J_{x}
    -\left[\frac{3}{8 \pi} \int 2 \beta_{x}\beta_{y}\frac{K_{3}}{B \rho} \, {\mathrm{d}}s \right] J_{y}
\end{eqnarray}
and
\begin{eqnarray}
    \Delta Q_{y} = \left[\frac{3}{8 \pi} \int \beta_{y}^{2}\frac{K_{3}}{B \rho} \, {\mathrm{d}}s \right] J_{y}
    -\left[\frac{3}{8 \pi} \int 2 \beta_{x}\beta_{y}\frac{K_{3}}{B \rho}\, {\mathrm{d}}s \right] J_{x}.
\end{eqnarray}


\begin{thebibliography}{99}
\bibitem{bib:langmuir} I. Langmuir and L. Tonks, \textit{Phys. Rev.} {\bf{33}} (1929) 195.
\bibitem{bib:landau1946} L.D. Landau, \textit{J. Phys. USSR} {\bf{10}} (1946)  26.
\bibitem{bib:bohm1} D. Bohm and E. Gross, \textit{Phys. Rev.} {\bf{75}} (1949) 1851.
\bibitem{bib:bohm2} D. Bohm and E. Gross, \textit{Phys. Rev.} {\bf{75}} (1949) 1864.
\bibitem{bib:sagan} D. Sagan, {{On the physics of Landau damping}}, CLNS~93/1185~(1993).
\bibitem{bib:case} K. Case, \textit{Ann. Phys.} {\bf{7}} (1959) 349.
\bibitem{bib:kampen} N.G. Van Kampen, \textit{Physica} {\bf{21}} (1955) 949.
\bibitem{bib:malmberg} J. Malmberg and C. Wharton, \textit{Phys. Rev. Lett.} {\bf{13}} (1964) 184.
\bibitem{bib:sessler1} V. Neil and A. Sessler, \textit{Rev. Sci. Instrum.} {\bf{36}} (1965) 429.
\bibitem{bib:sessler2} L. Laslett, V. Neil and A. Sessler, \textit{Rev. Sci. Instrum.} {\bf{36}} (1965) 436.
\bibitem{bib:vlasov} A.A. Vlasov, \textit{J. Phys. USSR} {\bf{9}} (1945) 25.
\bibitem{bib:chao} A. Chao, {\emph{Theory of Collective Beam Instabilities in Accelerators}} (Wiley,~New~York, 1993).
\bibitem{bib:hofmann} A. Hofmann, {{Landau damping}}, Proc. CERN Accelerator School (2009).
\bibitem{bib:chaotig} A.~Chao and M.~Tigner, {\emph {Handbook of Accelerator Physics and Engineering}}
(World Scientific, Singapore, 1998).
\bibitem{bib:rumolo} G. Rumolo, {{Beam instabilities}}, these proceedings, CERN Accelerator School (2013).
\bibitem{bib:herr} W. Herr, {{Mathematical and numerical methods for non-linear beam dynamics}}, these proceedings, CERN Accelerator School (2013).
\bibitem{bib:keil} E. Keil and W. Schnell, {{Concerning longitudinal stability in the ISR}}, CERN-ISR-TH-RH/69-48 (1969).
\bibitem{bib:herrvos} W. Herr and L. Vos, {{Tune distributions and effective tune spread from beam--beam interactions and the consequences for Landau damping in the LHC}}, LHC~Project~Note~316~(2003).
\bibitem{bib:herrbb} W. Herr, {{Beam--beam effects}}, Proc. CERN Accelerator School (2003).
\bibitem{bib:pieloni} T. Pieloni, {{Beam--beam effects}}, these proceedings, CERN Accelerator School (2013).
\bibitem{bib:buffat} X. Buffat, {{Consequences of missing collisions -- beam stability and Landau damping}}, Proc.
     ICFA Beam--Beam Workshop 2013, CERN (2013).
\bibitem{bib:alexahin1} Y. Alexahin, W. Herr \textit{et al.}, {{Coherent beam--beam effects}}, Proc. HEACC 2001, Tsukuba, Japan, 2001.
\bibitem{bib:alexahin2} Y. Alexahin, {{A study of the coherent beam--beam effect in the framework of the Vlasov perturbation theory}}, LHC~Project~Report~461~(2001).
\bibitem{bib:alexahin3} Y. Alexahin, \textit{Nucl.~Instrum.~Methods} {\bf{A480}} (2002) 235.
\bibitem{bib:rgo} J. Scott Berg and F. Ruggiero, {{Landau damping with two-dimensional tune spread}}, CERN~SL-AP-96-71~(AP)~(1996).
\end{thebibliography}
\end{document}